\documentclass[lettersize,journal]{IEEEtran}
\usepackage{amsmath,amsfonts,amssymb}
\usepackage{algorithmic}
\usepackage{array}
\usepackage[caption=false,font=normalsize,labelfont=sf,textfont=sf]{subfig}
\usepackage{textcomp,gensymb}
\usepackage{stfloats}
\usepackage{url}
\usepackage{verbatim}
\usepackage{graphicx}
\usepackage{booktabs}
\usepackage{xcolor}
\usepackage{multirow,rotating}	
\usepackage{color, colortbl}

\usepackage{cite}
\def\BibTeX{{\rm B\kern-.05em{\sc i\kern-.025em b}\kern-.08em
    T\kern-.1667em\lower.7ex\hbox{E}\kern-.125emX}}
\usepackage{balance}
\usepackage{tablefootnote}
\usepackage{refcount}
\usepackage{ulem}

\usepackage{array}
\newcolumntype{H}{>{\setbox0=\hbox\bgroup}c<{\egroup}@{}}

\definecolor{LightCyan}{rgb}{0.88,1,1}

\makeatletter
\def\ps@IEEEtitlepagestyle{%
  \def\@oddfoot{\mycopyrightnotice}%
  \def\@oddhead{\hbox{}\@IEEEheaderstyle\leftmark\hfil\thepage}\relax
  \def\@evenhead{\@IEEEheaderstyle\thepage\hfil\leftmark\hbox{}}\relax
  \def\@evenfoot{}%
}
\def\mycopyrightnotice{%
  \begin{minipage}{\textwidth}
  \centering \scriptsize
  Copyright~\copyright 2023 IEEE.  Personal use of this material is permitted.  Permission from IEEE must be obtained for all other uses, in any current or future media, including reprinting/republishing this material for advertising or promotional purposes, creating new collective works, for resale or redistribution to servers or lists, or reuse of any copyrighted component of this work in other works.
  \end{minipage}
}
\makeatother

\usepackage[acronym,shortcuts]{glossaries}
\newacronym{ASR}{ASR}{automatic speech recognition}
\newacronym{BSS}{BSS}{blind source separation}
\newacronym{BLSTM}{BLSTM}{bidirectional long short-term memory}
\newacronym{CNN}{CNN}{ convolutional neural network}
\newacronym{DNN}{DNN}{deep \gls{NN}}
\newacronym{DNS}{DNS}{deep noise suppression}
\newacronym{DOA}{DOA}{direction of arrival}
\newacronym{EEG}{EEG}{electroencephalogram}
\newacronym{fHMM}{F-HMM}{factorial hidden Markov model}
\newacronym{FiLM}{FiLM}{Feature-wise Linear Modulation}
\newacronym{GMM}{GMM}{Gaussian mixture model}
\newacronym{UBM}{UBM}{Universal Background model}
\newacronym{HMM}{HMM}{hidden Markov model}
\newacronym{ICA}{ICA}{independent component analysis}
\newacronym{ILD}{ILD}{interaural level difference}
\newacronym{IPD}{IPD}{interaural phase difference}
\newacronym{TPD}{TPD}{target phase difference}
\newacronym{iSTFT}{iSTFT}{inverse \gls{STFT}}
\newacronym{ILRMA}{ILRMA}{independent low-rank matrix analysis}
\newacronym{IVA}{IVA}{independent vector analysis}
\newacronym{IVE}{IVE}{independent vector extraction}
\newacronym{LRS}{LRS}{Lip Reading Sentences}
\newacronym{LSTM}{LSTM}{long short-term memory}
\newacronym{MSE}{MSE}{mean squared error}
\newacronym{MVDR}{MVDR}{minimum variance distortionless response}
\newacronym{NMF}{NMF}{non-negative matrix factorization}
\newacronym{NN}{NN}{neural network}
\newacronym{PESQ}{PESQ}{perceptual evaluation of speech quality}
\newacronym{PIT}{PIT}{permutation invariant training}
\newacronym{RNN}{RNN}{recurrent neural network}
\newacronym{SDR}{SDR}{signal-to-distortion ratio}
\newacronym{SI-SNR}{SI-SNR}{scale-invariant \gls{SNR}}
\newacronym{SNR}{SNR}{signal-to-noise ratio}
\newacronym{STFT}{STFT}{short-time Fourier transform}
\newacronym{STOI}{STOI}{short-time objective intelligibility}
\newacronym{TSASR}{TS-ASR}{target speaker \gls{ASR}}
\newacronym{TSE}{TSE}{target speech/speaker extraction}
\newacronym{TSVAD}{TS-VAD}{target speaker \gls{VAD}}
\newacronym{VAD}{VAD}{voice activity detection}
\newacronym{WER}{WER}{word error rate}
\newacronym{WHAM}{WHAM}{WSJ0 Hipster Ambient Mixtures}
\newacronym{WSJ}{WSJ}{Wall Street Journal}
\newacronym{VAE}{VAE}{variational autoencoder}

\newcommand{\tablefootnotemark}[1]{\textsuperscript{\getrefnumber{#1}}}

\newcommand\rev[1]{\textcolor{black}{#1}}

\DeclareMathOperator{\Diag}{diag}

\def\m{m} 
\def\t{t} 
\def\f{f} 
\def\n{n} 
\def\tgt{s} 

\def\N{N} 

\def\Zmix{\mathbf{Z}_\mix} 

\def\Ztgt{\mathbf{Z}_\tgt} 

\def\W{\mathbf{L}} 

\def\mix{y}
\def\mixm[#1]{\mix^\m[#1]}
\def\mixmvec{\mathbf{\mix}^m} 
\def\mixvec{\mathbf{\mix}} 
\def\mixf{Y}

\def\mixfmat{\mathbf{\mixf}} 

\def\src{x}
\def\srcmt[#1]{\src_{#1}^\m[\t]} 
\def\srcmvec[#1]{\mathbf{\src}_{#1}^m} 
\def\srct[#1]{\src_{#1}[t]} 
\def\srcvec[#1]{\mathbf{\src}_{#1}} 
\def\srcf{X}
\def\srcfmnf[#1]{\srcf_{#1}^\m[\n,\f]} 
\def\srcfmmat[#1]{\mathbf{\srcf}_{#1}^\m} 
\def\srcfnf[#1]{\srcf_{#1}[\n,\f]} 
\def\srcfmat[#1]{\mathbf{\srcf}_{#1}} 

\def\estsrc{\hat{\src}}
\def\estsrct[#1]{\estsrc_{#1}[t]} 
\def\estsrcvec[#1]{\hat{\mathbf{\src}}_{#1}}
\def\estsrcmt[#1]{\hat{\mathbf{\srcf}}_{#1}}
\def\estfmnf[#1]{\hat{S}_{#1}^\m[\n,\f]} 

\def\int{i}
\def\intmvec{\mathbf{\int}^m} 
\def\intvec{\mathbf{\int}} 

\def\nois{v}
\def\noismvec{\mathbf{\nois}^m} 
\def\noisvec{\mathbf{\nois}} 

\def\cluemat{\uppercase{\mathbf{C}}_\tgt}

\def\acluemat{\mathbf{C}_\tgt^{(a)}}

\def\vcluemat{\uppercase{\mathbf{C}}_\tgt^{(v)}}

\def\dcluemat{\mathbf{C}_\tgt^{(d)}}

\def\emb{\mathbf{E}_\tgt}
\def\aemb{\mathbf{E}_\tgt^{(a)}}
\def\vemb{\uppercase{\mathbf{E}}_\tgt^{(v)}}

\def\maskmt{\mathbf{M}_\tgt}

\def\BF{W}
\def\mixfnfmmat{\mathbf{\mixf}[\n,\f]} 

\def\estfnfm{\hat{\srcf}_\tgt[\n,\f]} 

\DeclareMathOperator*{\argmin}{arg\,min}

\DeclareMathOperator{\TSE}{TSE}
\DeclareMathOperator{\BSS}{BSS}
\DeclareMathOperator{\Denoise}{Denoise}

\DeclareMathOperator{\clueEnc}{ClueEncoder}

\DeclareMathOperator{\Extractor}{TgtExtractor}

\DeclareMathOperator{\MixEnc}{MixEncoder}
\DeclareMathOperator{\MixNet}{MixNet}
\DeclareMathOperator{\FE}{FE}
\DeclareMathOperator{\VFE}{VFE}
\DeclareMathOperator{\Fusion}{Fusion}

\DeclareMathOperator{\Reconstruct}{Reconstruct}
\DeclareMathOperator{\MaskNet}{MaskNet}

\DeclareMathOperator{\NN}{NN}
\DeclareMathOperator{\upsample}{Upsample}

\begin{document}
\title{Neural Target Speech Extraction: An Overview}
\author{\begin{tabular}{c}%
Katerina Zmolikova, %
Marc Delcroix,
Tsubasa Ochiai,
Keisuke Kinoshita, 
Jan \v{C}ernock{\'y},
Dong Yu
\thanks{Katerina Zmolikova and Jan \v{C}ernock{\'y} are with Brno University of Technology, Speech@FIT. Marc Delcroix, Tsubasa Ochiai and Keisuke Kinoshita are with NTT Corporation. Dong Yu is with Tencent, AI Lab.}
\end{tabular}
}


\maketitle


\begin{abstract}
Humans can listen to a target speaker even in challenging acoustic conditions that have noise, reverberation, and interfering speakers. This phenomenon is known as the cocktail-party effect. For decades, researchers have focused on approaching the listening ability of humans. One critical issue is handling interfering speakers because the target and non-target speech signals share similar characteristics, complicating their discrimination. Target speech/speaker extraction (TSE) isolates the speech signal of a target speaker from a mixture of several speakers with or without noises and reverberations using clues that identify the speaker in the mixture. Such clues might be a spatial clue indicating the direction of the target speaker, a video of the speaker's lips, or a pre-recorded enrollment utterance from which their voice characteristics can be derived. 
TSE is an emerging field of research that has received increased attention in recent years because it offers a practical approach to the cocktail-party problem and involves such aspects of signal processing as audio, visual, array processing, and deep learning. This paper focuses on recent neural-based approaches and presents an in-depth overview of TSE. We guide readers through the different major approaches, emphasizing the similarities among frameworks and discussing potential future directions.

\end{abstract}


\section{Introduction}

In everyday life, we are constantly immersed in complex acoustic scenes consisting of multiple sounds, such as a mixture of speech signals from multiple speakers and background noise from air-conditioners or music. Humans naturally extract relevant information from such noisy signals as they enter our ears. The cocktail-party problem is a typical example \cite{bronkhorst2015cocktail}, where we can follow the conversation of a speaker of interest (\emph{target speaker}) in a noisy room with multiple interfering speakers. Humans can manage this complex task due to \emph{selective attention} or a \emph{selective hearing} mechanism that allows us to focus our attention on a target speaker's voice and ignore others. Although the mechanisms of human selective hearing are not fully understood yet, many studies have identified essential cues exploited by humans to attend to a target speaker in a speech mixture: spatial, spectral (audio), visual, or semantic cues \cite{bronkhorst2015cocktail}. One long-lasting goal of speech processing research is designing machines that can achieve similar listening abilities as humans, i.e., selectively extracting the speech of a desired speaker based on auxiliary cues.

\begin{figure}[tb]
    \centering
    \includegraphics[width=1\linewidth]{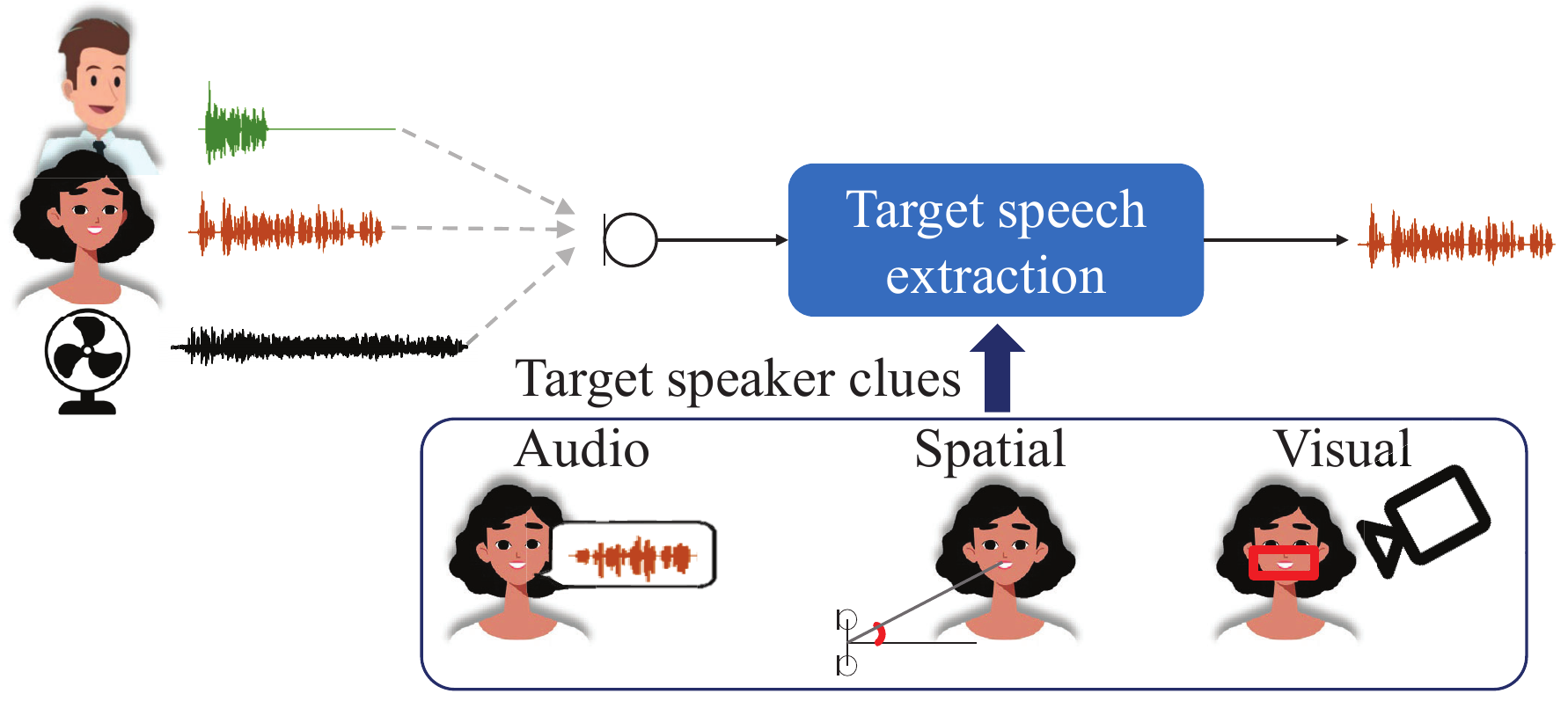}
    \caption{\gls{TSE} problem and examples of clues}
    \label{fig:problem}
\end{figure}

In this paper, we present an overview of recent developments in \rev{\gls{TSE}}, which estimates the speech signal of a target speaker in a mixture of several speakers, given auxiliary cues to identify the target\footnote{Alternative terms in the literature for \gls{TSE} include \rev{informed source separation,} personalized speech enhancement, or audio-visual speech separation, \rev{depending on the context and the modalities involved}.}. 
In the following, we call auxiliary cues, \emph{clues}, since they represent hints for identifying the target speaker in the mixture.
Fig.~\ref{fig:problem} illustrates the \gls{TSE} problem and shows that by exploiting the clues, \gls{TSE} can focus on the voice of the target speaker while ignoring other speakers or noise. Inspired by psychoacoustic studies\cite{bronkhorst2015cocktail}, several clues have been explored to tackle the \gls{TSE} problem, such as spatial clues that provide the direction of the target speaker \cite{Flanagan_85,gu2019neural}, visual clues from video of \rev{their} face \cite{hershey2001audio,rivet2014audiovisual,michelsanti2021overview,afouras2018conversation,ephrat2018looking,owens2018audio}, or audio clues extracted from pre-recorded enrollment recording of \rev{their} voice \cite{zmolikova2019speakerbeam,wang2018voicefilter,Jansky_20}.

The \gls{TSE} problem is directly related to human selective hearing, although we approach it from an engineering point of view and do not try to precisely mimic human mechanisms. 
\gls{TSE} is related to other speech and audio-processing tasks such as noise reduction and \gls{BSS} that do not use clues about the target speaker. Although noise reduction does suppress the background noise, it cannot handle well interfering speakers. \gls{BSS} estimates each speech source signal in a mixture, which usually requires estimating the number of sources, a step that is often challenging. Moreover, it estimates the source signals without identifying them, which leads to global permutation ambiguity at its output; it remains ambiguous which of the estimated source signals corresponds to the target speaker. In contrast, \gls{TSE} focuses on the target speaker's speech by exploiting clues without assuming knowledge of the number of speakers in the mixture and avoids global permutation ambiguity. It thus offers a practical alternative to noise reduction or \gls{BSS} when the use case requires focusing on a desired speaker’s voice.

Solving the \gls{TSE} problem promises real implications for the development of many applications: (1)~robust voice user interfaces or voice-controlled smart devices that only respond to a specific user; (2)~teleconferencing systems that can remove interfering speakers close by; (3)~hearing aids/hearables that can emphasize the voice of a desired interlocutor.

\gls{TSE} ideas can be traced back to early works on beamformers\cite{Flanagan_85}. Several works also extended \gls{BSS} approaches to exploit clues about the target speaker\cite{hershey2001audio,rivet2014audiovisual,Jansky_20}. Most of these approaches required a microphone array \cite{rivet2014audiovisual} or models trained on a relatively large amount of speech data from the target speaker \cite{hershey2001audio}. The introduction of \glspl{NN} enabled the building of powerful models that learn to perform complex conditioning on various clues by leveraging large amounts of speech data of various speakers. This evolution resulted in impressive extraction performance. Moreover, neural \gls{TSE} systems can operate with a single microphone and with speakers unseen during the training of the models, allowing more flexibility.  

This overview paper covers recent \gls{TSE} development and focuses on neural approaches. 
Its remaining sections are organized as follows. In Section \ref{sec:problem}, we formalize the \gls{TSE} problem and its relation to noise reduction and \gls{BSS} and introduce its historical context.
We then present a taxonomy of \gls{TSE} approaches and motivate the focus of this overview paper in Section \ref{sec:taxonomy}. We describe a general neural \gls{TSE} framework in Section \ref{sec:general_framework}. The later sections (\ref{sec:audio_tse}, \ref{sec:visual_tse}, and \ref{sec:spatial_tse}) introduce implementations of \gls{TSE} with different clues, such as audio, visual, and spatial clues. We discuss extensions to other tasks in Section \ref{sec:extension}. Finally, we conclude by describing the outlook on remaining issues in Section \ref{sec:outlook} and provide pointers to available resources for experimenting with \gls{TSE} in Section \ref{sec:resources}.

\section{Problem definition}
\label{sec:problem}
\subsection{Speech recorded with a distant microphone}
Imagine recording a target speaker's voice in a living room using a microphone placed on a table. This scenario represents a typical use case of a voice-controlled smart device or a video-conferencing device in a remote-work situation. Many sounds may co-occur while the speaker is speaking, e.g., a vacuum cleaner, music, children screaming, voices from another conversation, or from a TV.
The speech signal captured at a microphone thus consists of a mixture of the target speaker's speech and interference from the speech of other speakers and background noise\footnote{In this paper, we do not explicitly consider the effect of reverberation caused by the reflection of sounds on the walls and surfaces in a room, which also corrupt the recorded signal. Some of the approaches we discussed implicitly handle reverberation.}. We can express the mixture signal recorded at a microphone as
\begin{align}
    \mixmvec = \srcmvec[\tgt] + \underbrace{\sum_{k\neq s} \srcmvec[k] + \noismvec}_{\triangleq \intmvec},
    \label{eq:mixt_1ch_time}
\end{align}
where $\mixmvec = [\mixm[0], \ldots, \mixm[T] ] \in \mathbb{R}^T$, $\srcmvec[\tgt] \in \mathbb{R}^T$,  $\srcmvec[k] \in \mathbb{R}^T$, and $\noismvec \in \mathbb{R}^T$ are the time-domain signal of the mixture, the target speech, the interference speech, and noise signals, respectively. \rev{Variable} $T$ represents the duration (number of samples) of the signals, $m$ is the index of the microphone in an array of microphones, $s$ represents the index of the target speaker and $k$ is the index for the other speech sources. We drop microphone index $m$ whenever we deal with single microphone approaches. In the \gls{TSE} problem, we are interested in only recovering the target speech of speaker $s$, $\srcmvec[\tgt]$, and view all the other sources as undesired signals to be suppressed. We can thus define the interference signal as $\intmvec \in \mathbb{R}^T$. Note that we make no explicit hypotheses about the number of interfering speakers.

\subsection{TSE problem and its relation to BSS and noise reduction}
The \gls{TSE} problem is to estimate the target speech, given \rev{a clue, $\cluemat$, as}
\begin{align}
    \estsrcvec[s]  = \TSE(\mixvec, \cluemat; \theta^{\text{TSE}}),
    \label{eq:tse}
\end{align}
where  $\estsrcvec[s]$ is the estimate of the target speech, $\TSE(\cdot; \theta^{\text{TSE}})$ represents a \gls{TSE} system with parameters $\theta^{\text{TSE}}$. 
\rev{The clue, $\cluemat$, allows identifying the target speaker in the mixture. It} can be of various types, such as a pre-recorded enrollment utterance, $\acluemat$, a video signal capturing the face or lips movements of the target speaker, $\vcluemat$, or such spatial information as the \gls{DOA} of the speech of the target speaker, $\dcluemat$. 

\begin{figure*}[tb]
    \centering
    \includegraphics[width=0.9\linewidth]{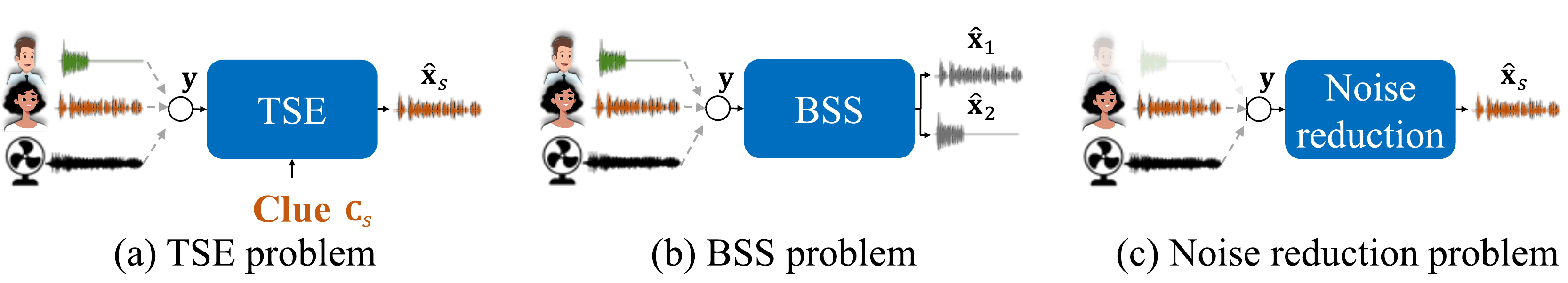}
    \caption{Comparison of TSE with \gls{BSS} and noise reduction}
    \label{fig:TSE_SEP_Denoising}
\end{figure*}
In the later sections, we expand on how to design \gls{TSE} systems. Here, we first emphasize the key difference between \gls{TSE} and \gls{BSS} and noise reduction. Fig.~\ref{fig:TSE_SEP_Denoising} compares these three problems.

\gls{BSS}\cite{sawada2019review,wang2018supervised} estimates all the source signals in a mixture without requiring clues:
\begin{align}
    \{\estsrcvec[1], \ldots, \estsrcvec[K]\}  = \BSS(\mixvec; \theta^{\text{BSS}}),
    \label{eq:bss}
\end{align}
where $ \BSS(\cdot;\theta^{\text{BSS}})$ represents a separation system with parameters $\theta^{\text{BSS}}$, $\estsrcvec[k]$ are the estimates of the speech sources, and $K$ is the number of sources in the mixture. As seen in Eq.~\eqref{eq:bss}, \gls{BSS} does not and cannot differentiate the target speech from other speech sources. Therefore, we cannot know in advance which output corresponds to the target speech, i.e., there is a \emph{global permutation ambiguity} problem between the outputs and the speakers. Besides, since the number of outputs is given by the number of sources, the number of sources $K$ must be known or estimated. Comparing Eqs. \eqref{eq:tse} and \eqref{eq:bss} emphasizes the fundamental difference between \gls{TSE} and \gls{BSS}: (1)~\gls{TSE} estimates only the target speech signal, while \gls{BSS} estimates all the signals, and (2)~\gls{TSE} is conditioned on speaker clue $\cluemat$, while \gls{BSS} only relies on the observed mixture\footnote{Another setup sitting between \gls{TSE} and \gls{BSS} is a task that extracts multiple target speakers, e.g., extracting the speech of all the meeting attendees given such information about them as enrollment or videos of all the speakers.}. Typical use cases for \gls{BSS} include applications that require estimating speech signals of every speaker, such as automatic meeting transcription systems.

Noise reduction is another related problem. It assumes that the interference only consists of background noise, i.e., $\intvec = \noisvec$, and can thus enhance the target speech without requiring clues:
\begin{align}
    \estsrcvec[s]  = \Denoise(\mixvec; \theta^{\text{Denoise}}),
    \label{eq:denoise}
\end{align}
where $\Denoise(\cdot; \theta^{\text{Denoise}})$ represents a noise reduction system with parameters $\theta^{\text{Denoise}}$. Unlike \gls{BSS}, a noise reduction system’s output only consists of target speech $\estsrcvec[s]$, and there is thus no global permutation ambiguity. This is possible if the background noise and speech have distinct characteristics. For example, we can assume that ambient noise and speech signals exhibit different spectro-temporal characteristics that enable their discrimination.
However, noise reduction cannot suppress interfering speakers because it cannot discriminate among different speakers in a mixture without clues\footnote{Some works propose to exploit clues for noise reduction and apply similar ideas of \gls{TSE} to reduce  background noise (and sometimes interfering speakers). In the literature, this is called personalized speech enhancement, which in this paper, we view as a special case of the \gls{TSE} problem, where only the target speaker is actively speaking\cite{Eskimez2022personalized}.}.
Noise reduction is often used, e.g., in video-conferencing systems or hearing aids.

\gls{TSE} is an alternative to \gls{BSS} and noise reduction, which uses a clue to simplify the problem. 
Like \gls{BSS}, it can handle speech mixtures. Like noise reduction, it only estimates the target speaker, thus avoiding global permutation ambiguity and the need to estimate the number of sources. However, \gls{TSE} requires access to clues, unlike \gls{BSS} and noise reduction. 
Moreover, it must internally perform two sub-tasks: (1)~identifying the target speaker and (2)~estimating the speech of that speaker in the mixture. 
\gls{TSE} is thus a challenging problem that introduces specific issues and requires dedicated solutions. 

A straightforward way to achieve \gls{TSE} using \gls{BSS} methods is to first apply \gls{BSS} and next select the target speaker among the estimated sources. Such a cascade system allows the separate development of \gls{BSS} and speaker identification modules. However, this scheme is usually computationally more expensive and imports some disadvantages of \gls{BSS}, such as the need to estimate the number of speakers in the mixture. Therefore, we focus on approaches that directly exploit the clues in the extraction process.  Nevertheless, most \gls{TSE} research is rooted in \gls{BSS}, as argued in the following discussion on the historical context.

\subsection{Historical context}
The first studies related to \gls{TSE} were performed in the 1980s. Flanagan et al.\cite{Flanagan_85} explored enhancing a target speaker's voice in a speech mixture, assuming that the target speech originated from a fixed and known direction. They employed a microphone array to record speech and designed a fixed beamformer that enhanced the signals from the target direction\cite{Flanagan_85,gannot2017consolidated}.  We consider that this work represents an early \gls{TSE} system that relies on spatial clues.

In the mid-1990s, the \gls{BSS} problem gained attention with pioneering works on \gls{ICA}. \gls{ICA} estimates spatial filters that separate the sources by relying on the assumption of the independence of the sources in the mixture and the fact that speech signals are non-Gaussian\cite{sawada2019review}. A frequency-domain \gls{ICA} suffers from a frequency permutation problem because it treats each frequency independently. In the mid-2000s, \gls{IVA} addressed the frequency-permutation problem by working on vectors spanning all frequency bins, which allowed modeling dependency among frequencies\cite{sawada2019review}. Several works have extended \gls{ICA} and \gls{IVA} to perform \gls{TSE}, which simplifies inference by focusing on a single target source. For example, in the late 2000s, \gls{TSE} systems were designed by incorporating the voice activity information of the target speaker derived from video signals to the \gls{ICA} criterion, allowing identification and extraction of only the target source\cite{rivet2014audiovisual}. In the late 2010s, \gls{IVE} extended \gls{IVA} to extract a single source out of the mixture. In particular, \gls{IVE} exploits clues to guide the extraction process, such as the enrollment of the target speaker to achieve \gls{TSE} \cite{Jansky_20}. All these approaches require a microphone array to capture speech.

In the first decade of the 2000s, single-channel approaches for \gls{BSS} emerged, such as \gls{fHMM}\cite{Hershey2010Super} and \gls{NMF}\cite{virtanen2007monaural}. These approaches relied on pre-trained spectral models of speech signals learned on clean speech data. 
A\rev{n} \gls{fHMM} is a model of speech mixtures, where the speech of each speaker in the mixture is explicitly \rev{modeled} using a separate \gls{HMM}. The parameters of each speaker-\gls{HMM} are learned on the clean speech data of that speaker. The separation process involves inferring the most likely \gls{HMM} state sequence associated with each speaker-\gls{HMM}, which requires approximations to make inference tractable.
This approach was the first to achieve super-human performance using only single-channel speech \cite{Hershey2010Super}. In the early 2000s, \gls{fHMM} was also among the first approaches to exploit visual clues\cite{hershey2001audio}\footnote{This framework needs having clues for all of the speakers, a requirement that negates some of the advantages of \gls{TSE}, e.g., the number of speakers must be known beforehand. Despite that, the method does not suffer from global permutation ambiguity, since visual clues identify the target speaker, and we thus include this work in the broader view of \gls{TSE} methods.}. In \gls{NMF}, the spectrogram of each source is modeled as a multiplication of pre-learned bases, representing the basic spectral patterns and their time-varying activations. \gls{NMF} methods have also been extended to multi-channel signals \cite{sawada2019review} and used to extract a target speaker \cite{ozerov2011using} with a flexible multi-source model of the background. The main shortcoming of the \gls{fHMM} and \gls{NMF} methods is that they require pre-trained source models and thus struggle with unseen speakers. Furthermore, the inference employs a computationally expensive iterative optimization.

In the mid-2010s, \glspl{DNN} were first introduced to address the \gls{BSS} problem. These approaches rapidly gained attention with the success of deep-clustering and \gls{PIT} \rev{\cite{hershey2016deep,yu2016permutation}}, which showed that single-channel speaker-open\footnote{\gls{BSS} is possible for speakers unseen during training, i.e., not present in the training data.} \gls{BSS} was possible. In particular, the introduction of \glspl{DNN} enabled more accurate and flexible spectrum modeling and computationally efficient inference. These advances were facilitated by supervised training methods that can exploit a large amount of data. 

Neural \gls{BSS} rapidly influenced \gls{TSE} research. For example, Du et al. \cite{du2014speech} trained a speaker-close \gls{NN} to extract the speech of a target speaker using training data with mixed various interfering speakers. This work is an initial neural \gls{TSE} system using audio clues. However, using speaker-close models requires a significant amount of data from the target speaker and cannot be extended to speakers unseen during training. 
Subsequently, the introduction of \gls{TSE} systems conditioned on speaker characteristics derived from an enrollment utterance significantly mitigated this requirement\cite{zmolikova2019speakerbeam,wang2018voicefilter,xu2019time}. Enrollment consists of a recording of a target speaker's voice, which amounts to a few seconds of speech. With these approaches, audio clue-based \gls{TSE} became possible for speakers unseen during training as long as an enrollment utterance was available. 
Furthermore, the flexibility of \glspl{NN} to integrate different modalities combined with the high modeling capability of face recognition or lip-reading systems offered new possibilities for speaker-open visual clue-based \gls{TSE}\cite{ephrat2018looking,afouras2018conversation}. More recently, neural approaches have also been introduced for spatial-clue-based \gls{TSE} \cite{gu2019neural,heitkaemper2019study}. 

\gls{TSE} has gained increased attention. For example, dedicated tasks were part of such recent evaluation campaigns as the \gls{DNS}\footnote{\url{ https://www.microsoft.com/en-us/research/academic-program/deep-noise-suppression-challenge-icassp-2022/ }} and Clarity\footnote{\url{https://claritychallenge.github.io/clarity_CC_doc}} challenges. Many works have extended \gls{TSE} to other tasks, such as a direct \gls{ASR} of a target speaker from a mixture, which is called \gls{TSASR} \cite{delcroix2018single,denisov2019end}, or personalized \gls{VAD}/diarization \cite{ding2019personal,medennikov2020target}. Notably, \gls{TSVAD}-based diarization\cite{medennikov2020target} has been very successful in such evaluation campaigns as CHiME-6\footnote{\url{https://chimechallenge.github.io/chime6/results.html}} or DIHARD-3\footnote{\url{https://dihardchallenge.github.io/dihard3/results}}, outperforming state-of-the-art diarization approaches in challenging conditions.


\section{TSE Taxonomy }
\label{sec:taxonomy}

\begin{table*}[tb]
    \centering
    \caption{Taxonomy of \gls{TSE} works: Approaches within scope of this overview paper are emphasized in red.}
    \begin{tabular}{@{}llcc ccc cc cc}
    \toprule
         & Representative approaches & References & Year &\multicolumn{3}{c}{Type of clues} &  \multicolumn{2}{c}{Number of mic.} &  \multicolumn{2}{c}{Speaker-close/open} \\
         &&&& Audio & Visual & Spatial & Single & Multi & Close & Open  \\
         \midrule
         
         & Fixed beamforming &\cite{Flanagan_85,gannot2017consolidated}\footnotemark& 1985 &- & - & \checkmark & - & \checkmark & - & \checkmark \\
         \midrule
         \multirow{5}{*}{\rotatebox[origin=c]{90}{Generative}}& Audio-visual \gls{fHMM} & \cite{hershey2001audio} & 2001 &\checkmark\footnotemark & \checkmark & - & \checkmark & - & \checkmark & -  \\
         & ICA with visual voice activity &\cite{rivet2014audiovisual} & 2007 & - & \checkmark & - & - & \checkmark &  - & \checkmark\\
         & Multi-channel \gls{NMF} & \cite{ozerov2011using} & 2011&\checkmark\tablefootnotemark{spkclose} & - & - & - & \checkmark  & \checkmark & -  \\
         & \gls{IVE} with x-vectors & \cite{Jansky_20} & 2020&\checkmark & - & - & - & \checkmark & - & \checkmark  \\
         & Audio-visual VAE & \cite{sadeghu2020audovisual} & 2020& - & \checkmark & - & \checkmark & - & - & \checkmark  \\
         \midrule
         \multirow{15}{*}{\rotatebox[origin=c]{90}{Discriminative}}
         & Speaker-specific network & \cite{du2014speech} & 2014&\checkmark\tablefootnotemark{spkclose} & - & - & \checkmark & - &  \checkmark & -  \\
           & \cellcolor{red!25}Multi-channel SpeakerBeam &\cellcolor{red!25}  \cite{zmolikova2017speaker,zmolikova2019speakerbeam} &\cellcolor{red!25} 2017 & \cellcolor{red!25}\checkmark & \cellcolor{red!25} - &  \cellcolor{red!25}- &\cellcolor{red!25} - & \cellcolor{red!25}\checkmark &\cellcolor{red!25} - &\cellcolor{red!25}\checkmark  \\
         & \cellcolor{red!25}SpeakerBeam &\cellcolor{red!25} \cite{zmolikova2019speakerbeam} &\cellcolor{red!25} 2019 & \cellcolor{red!25}  \checkmark &\cellcolor{red!25} - &\cellcolor{red!25}  - &\cellcolor{red!25} \checkmark &\cellcolor{red!25} - &\cellcolor{red!25} - &\cellcolor{red!25}\checkmark  \\
         & \cellcolor{red!25}VoiceFilter &\cellcolor{red!25} \cite{wang2018voicefilter} &\cellcolor{red!25} 2019 & \cellcolor{red!25}  \checkmark &\cellcolor{red!25} - &\cellcolor{red!25}  - &\cellcolor{red!25} \checkmark &\cellcolor{red!25} - &\cellcolor{red!25} - &\cellcolor{red!25}\checkmark  \\
         & \cellcolor{red!25}SpEx  &\cellcolor{red!25} \cite{ge2020spex} &\cellcolor{red!25} 2020 & \cellcolor{red!25}  \checkmark &\cellcolor{red!25} - &\cellcolor{red!25}  - &\cellcolor{red!25} \checkmark &\cellcolor{red!25} - &\cellcolor{red!25} - &\cellcolor{red!25}\checkmark \\
         &\cellcolor{red!25}The conversation &\cellcolor{red!25} \cite{afouras2018conversation} &\cellcolor{red!25} 2018 & \cellcolor{red!25} - &\cellcolor{red!25} \checkmark &\cellcolor{red!25} - &\cellcolor{red!25} \checkmark &\cellcolor{red!25} - & \cellcolor{red!25} - &\cellcolor{red!25} \checkmark \\
         &\cellcolor{red!25}Looking-to-listen &\cellcolor{red!25} \cite{ephrat2018looking} &\cellcolor{red!25} 2018 & \cellcolor{red!25} - &\cellcolor{red!25} \checkmark &\cellcolor{red!25} - &\cellcolor{red!25} \checkmark &\cellcolor{red!25} - &\cellcolor{red!25} - &\cellcolor{red!25} \checkmark  \\
         &\cellcolor{red!25}On/off-screen audio-visual separation &\cellcolor{red!25} \cite{owens2018audio} &\cellcolor{red!25} 2018 & \cellcolor{red!25} - &\cellcolor{red!25} \checkmark &\cellcolor{red!25} - &\cellcolor{red!25} \checkmark &\cellcolor{red!25} - & \cellcolor{red!25} - &\cellcolor{red!25} \checkmark \\
         &\cellcolor{red!25}Landmark-based AV speech enh. &\cellcolor{red!25} \cite{morrone2019face} &\cellcolor{red!25} 2019 & \cellcolor{red!25} - &\cellcolor{red!25} \checkmark &\cellcolor{red!25} - &\cellcolor{red!25} \checkmark &\cellcolor{red!25} - &\cellcolor{red!25} - &\cellcolor{red!25} \checkmark  \\
         &\cellcolor{red!25}Multi-modal SpeakerBeam &\cellcolor{red!25} \cite{ochiai2019multimodal,sato2021multimodal} &\cellcolor{red!25} 2019 & \cellcolor{red!25} \checkmark &\cellcolor{red!25} \checkmark &\cellcolor{red!25} - &\cellcolor{red!25} \checkmark &\cellcolor{red!25} - &\cellcolor{red!25} - &\cellcolor{red!25} \checkmark \\
         &\cellcolor{red!25}AV speech enh. through obstructions  &\cellcolor{red!25} \cite{afouras2019my} &\cellcolor{red!25} 2019 & \cellcolor{red!25} \checkmark &\cellcolor{red!25} \checkmark &\cellcolor{red!25} - &\cellcolor{red!25} \checkmark &\cellcolor{red!25} - &\cellcolor{red!25} - &\cellcolor{red!25} \checkmark \\
         &\cellcolor{red!25}Neural spatial filter&\cellcolor{red!25} \cite{gu2019neural} &\cellcolor{red!25} 2019 & \cellcolor{red!25} \checkmark &\cellcolor{red!25} -&\cellcolor{red!25} \checkmark &\cellcolor{red!25} - &\cellcolor{red!25} \checkmark &\cellcolor{red!25} - &\cellcolor{red!25} \checkmark \\
         &\cellcolor{red!25}Spatial speaker extractor &\cellcolor{red!25} \cite{heitkaemper2019study} &\cellcolor{red!25} 2019 & \cellcolor{red!25} \checkmark &\cellcolor{red!25} -&\cellcolor{red!25} \checkmark &\cellcolor{red!25} - &\cellcolor{red!25} \checkmark &\cellcolor{red!25} - &\cellcolor{red!25} \checkmark \\
         &\cellcolor{red!25}Multi-channel multi-modal \gls{TSE} &\cellcolor{red!25} \cite{gu2020multi} &\cellcolor{red!25} 2020 & \cellcolor{red!25} \checkmark &\cellcolor{red!25} \checkmark &\cellcolor{red!25} \checkmark &\cellcolor{red!25} - &\cellcolor{red!25} \checkmark &\cellcolor{red!25} - &\cellcolor{red!25} \checkmark \\
 \bottomrule
    \end{tabular}
    \label{tab:taxonomy}
\end{table*}

\gls{TSE} is a vast research area spanning a multitude of approaches. This section organizes them to emphasize their relations and differences. We categorized the techniques using four criteria: 1) type of clues, 2) number of channels, 3) speaker-close vs. open, and 4) generative vs. discriminative.
Table~\ref{tab:taxonomy} summarizes the taxonomy; the works in the scope of this overview paper are emphasized in red.

\subsection{Type of clue}

The type of clue used to determine the target speaker is an important factor in distinguishing among \gls{TSE} approaches. The most prominent types are audio, visual, and spatial clues. This classification also defines the main organization of this article, which covers such approaches in Sections \ref{sec:audio_tse}, \ref{sec:visual_tse}, and \ref{sec:spatial_tse}. Other types have and could be proposed, as we briefly discuss in Section \ref{sec:outlook}.

An \textit{audio clue} consists of a recording of a speech signal of the target speaker. Such a clue can be helpful, e.g., in the use case of personal devices, where the user can pre-record an example of \rev{their} voice. Alternatively, for long recordings, such as meetings, clues can be obtained directly from part of the recording. The interest in audio clues sharply increased recently with the usage of neural models for \gls{TSE}\cite{zmolikova2019speakerbeam,wang2018voicefilter,Jansky_20}.
Audio clues are perhaps the most universal, because they do not require using any additional devices, such as multiple microphones or a camera. However, the performance may be limited compared to other clues, since discriminating speakers based only on their voice characteristics is prone to errors due to inter- and intra-speaker variability. For example, the voice characteristics of different speakers, such as family members, often closely resemble each other. On the other hand, the voice characteristics of one speaker may change depending on such factors as emotions, health, or age.

A \textit{visual clue} consists of a video of the target speaker talking. This type is often constrained to the speaker’s face, sometimes just to the lip area. Unlike audio clues, visual clues are typically synchronized with audio signals that are processed, i.e., not pre-recorded. A few works also explored just using a photo of the speaker\rev{\cite{chung2020facefilter}}. 
Visual clues have been employed to infer the activity pattern and location of the target speaker \cite{rivet2014audiovisual} or to jointly model audio and visual signals \cite{rivet2014audiovisual,hershey2001audio}. Recent works usually use visual clues to guide discriminative models toward extracting the target speaker \cite{ephrat2018looking,afouras2018conversation,owens2018audio}.
Visual clues are especially useful when speakers in the recording have similar voices\rev{\cite{ephrat2018looking}}. However, they might be sensitive to physical obstructions of the speaker in the video.

A \textit{spatial clue} refers to the target speaker's location, e.g., the angle from the recording devices. The location can be inferred in practice from a video of the room or a recording of a speaker in the same position.  Extracting the speaker based on \rev{their} location has been researched from mid 1980’s, with beamforming techniques that pioneered this topic \cite{Flanagan_85,gannot2017consolidated}. More recent \gls{IVE} models use location for initialization \cite{Jansky_20}. Finally, several works have shown that \glspl{NN} informed by location can also achieve promising performance \cite{gu2019neural,heitkaemper2019study}. Spatial clues are inherently applicable only when a recording from multiple microphones is available. However, they can identify the target speaker in the mixture rather reliably, especially when the speakers are stationary.

Different clues may work better in different situations. For example, the performance with audio clues might depend on the similarity of voices of the present speakers, and obstructions in the video may influence visual clues. As such, it is advantageous to use multiple clues simultaneously to combine their strengths. Many works have combined audio and visual clues \cite{ochiai2019multimodal,hershey2001audio}, and some have even added spatial clues\cite{gu2020multi}.

\subsection{Number of microphones}
\addtocounter{footnote}{-1}\footnotetext{Since the first works that proposed beamforming were not model-based, we consider them neither generative nor discriminative.\label{beamforming}}
\addtocounter{footnote}{+1}\footnotetext{In speaker-close cases, the models are trained on target speaker’s audio. We consider this an audio clue in Table~\ref{tab:taxonomy}.\label{spkclose}}

Another way to categorize the \gls{TSE} approaches is based on the number of microphones (channels) they use. Multiple channels allow the spatial diversity of the sources to be exploited to help discriminate the target speaker from interference. Such an approach also closely follows human audition, where binaural signals are crucial for solving the cocktail-party problem. 

All approaches with spatial clues require using a microphone array to capture the direction information of the sources in the mixture\cite{Flanagan_85,gannot2017consolidated, gu2019neural,heitkaemper2019study,gu2020multi}. Some \gls{TSE} approaches that exploit audio or visual clues also assume multi-channel recordings, such as the extensions of \gls{ICA}/\gls{IVA} approaches\cite{rivet2014audiovisual,Jansky_20}. 

\rev{Multi-channel approaches generally generate extracted signals with better quality and are thus preferable when recordings from a microphone array are available. However, sometimes they might fail when the sources are located in the same direction from the viewpoint of the recording device. Moreover, adopting a microphone array is not always an option when developing applications due to cost restrictions. In such cases, single-channel approaches are requested.}
They rely on spectral models of speech mixture using either \gls{fHMM} or recently \glspl{NN} and exploit audio \cite{zmolikova2019speakerbeam,wang2018voicefilter}  or visual clues \cite{afouras2018conversation,ephrat2018looking} to identify the target speech.

Recent single-channel neural \gls{TSE} systems have achieved remarkable performance. Interestingly, such approaches can also be easily extended to multi-channel processing by augmenting the input with spatial features \cite{gu2019neural} or combining the processing with beamforming\cite{zmolikova2017speaker,heitkaemper2019study}, as discussed in Section \ref{sec:tse_beamformer}. For example, using a beamformer usually extracts a higher quality signal due to employing a spatial linear filter to perform extraction, which can benefit \gls{ASR} applications \cite{zmolikova2019speakerbeam}.

\subsection{Speaker-open vs speaker-close methods}
We usually understand the clues used by \gls{TSE} as short evidence about the target speaker obtained at the time of executing the method, e.g., one utterance spoken by the target speaker, a video of him/her speaking, or \rev{their} current location. There are, however, also methods that use a more significant amount of data from the target speaker (e.g., several hours of \rev{their} speech) to build a model specific to that person. These methods can also be seen as \gls{TSE} except that the clues involve much more data.

We refer to these two categories as the speaker-open and speaker-close methods\rev{\footnote{\rev{Speaker-open and speaker-close categories are sometimes referred to as speaker-independent and speaker-dependent, respectively. We avoid this terminology, as in \gls{TSE}, all systems are informed about the target speaker, and therefore the term speaker-independent might be misleading.}}}. In speaker-open methods, the data of the target speaker are available only during the test time, i.e., the model is trained on the data of different speakers. In contrast, the target speaker is part of the training data in speaker-close methods. Many methods in the past were speaker-close, e.g., \cite{hershey2001audio} or \cite{ozerov2011using}, where the models were trained on the clean utterances of the target speaker. Also, the first neural models for \gls{TSE} used a speaker-specific network \cite{du2014speech}. Most recent works on neural methods, which use a clue as an additional input, are speaker-open methods\cite{wang2018voicefilter,zmolikova2019speakerbeam,ephrat2018looking,afouras2018conversation,gu2019neural}. Recent \gls{IVE} methods \cite{Jansky_20} are also speaker-open, i.e., they guide the inference of \gls{IVE} using the embedding of a previously unseen speaker. 

\subsection{Generative vs discriminative}
We can classify \gls{TSE} into approaches using generative or discriminative models.

Generative approaches model the joint distribution of the observations, target signals, and clues. The estimated target speech is obtained by maximizing the likelihood. In contrast, discriminative approaches directly estimate the target speech signal given observations and clues.

In the \gls{TSE} literature, generative models were the dominant choice in the pioneering works, including one \cite{hershey2001audio} that used \glspl{HMM} to jointly model audio and visual modalities. \gls{IVE} \cite{Jansky_20} is also based on a generative model of the mixtures. 

The popularity of discriminative models, in particular \glspl{NN}, has increased since mid-2010’s, and such models today are the choice for many problems, including \gls{TSE}. With discriminative models, \gls{TSE} is treated as a supervised problem, where the parameters of a \gls{TSE} model are learned using artificially generated training data. The modeling power of \glspl{NN} enables us to exploit large amounts of such data to build strong speech models. Moreover, the versatility of \glspl{NN} enables complex dependencies to be learned between different types of observations (e.g., speech mixture and video/speaker embeddings), which allows the successful
conditioning of the extraction process on various clues. 
However, \glspl{NN} also bring new challenges, such as generalization to unseen conditions or high computational requirements \rev{\cite{maciejewski2019analysis}}. 

\rev{Some recent works have also explored using generative \glspl{NN}, such as \glspl{VAE} \cite{sadeghu2020audovisual}, which might represent a middle-ground between the traditional generative approaches and those using discriminative \glspl{NN}.}

\subsection{Scope of overview paper}
In the remainder of our paper, we focus on the neural methods for \gls{TSE} emphasized in Table \ref{tab:taxonomy}. 
Recent neural \gls{TSE} approaches opened the possibility of achieving high-performance extraction with various clues. They can be operated with a single microphone and applied for speaker-open conditions, which are very challenging constraints for other schemes. Consequently, these approaches have received increased attention from both academia and industry.

In the next section, we introduce a general framework to provide a uniformized view of the various \gls{NN}-based \gls{TSE} approaches, for both single- and multi-channel approaches, and independently of the type of clues.  
We then respectively review the approaches relying on audio, visual, and spatial clues in Sections \ref{sec:audio_tse}, \ref{sec:visual_tse}, and \ref{sec:spatial_tse}.

\section{General framework for neural TSE}
\label{sec:general_framework}
\begin{figure*}[tb]
    \centering
    \includegraphics[width=0.7\linewidth]{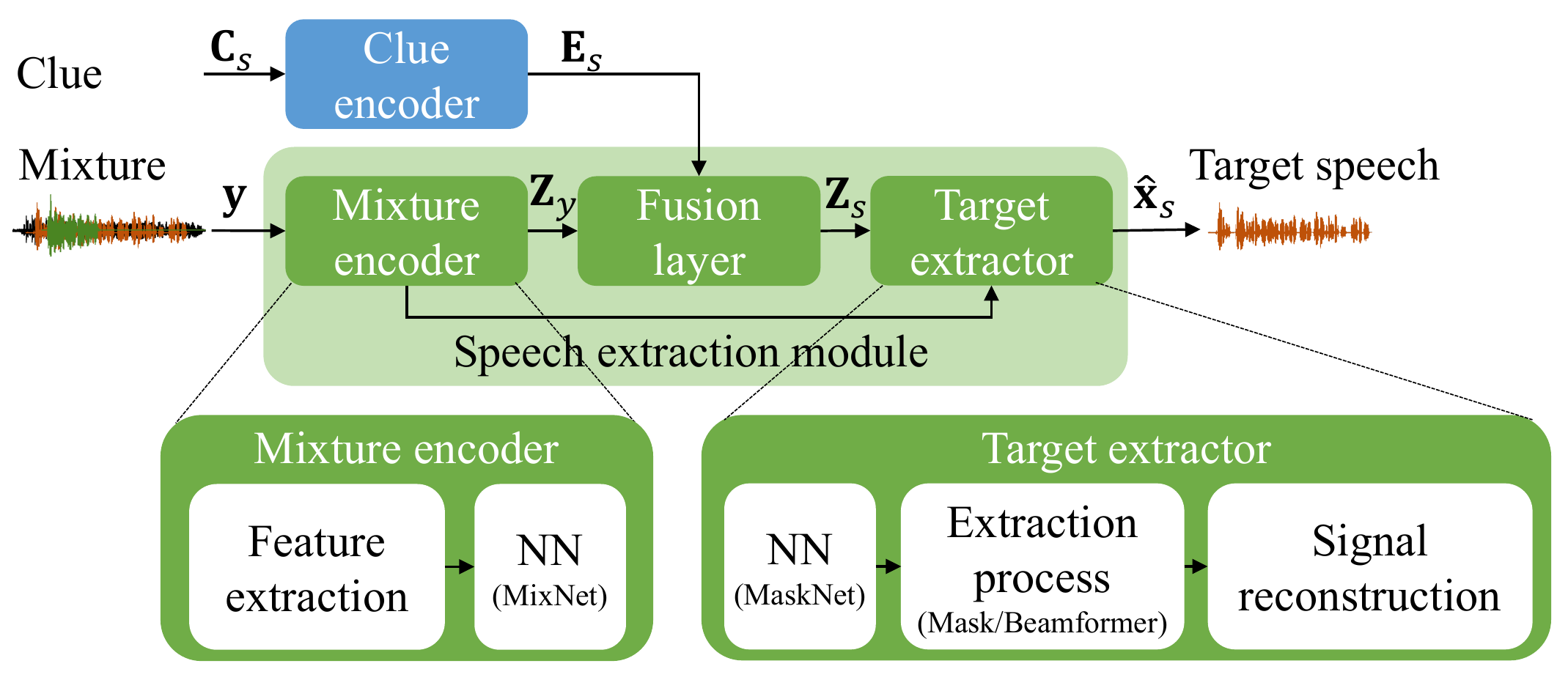}
    \caption{General  framework for neural \gls{TSE}}
    \label{fig:general_framework}
\end{figure*}

In the previous section, we introduced a taxonomy that described the diversity of approaches to tackle the \gls{TSE} problem. However, recent neural \gls{TSE} systems have much in common. In this section, we introduce a general framework that provides a unified view of a neural \gls{TSE} system, which shares the same processing flow independently of the type of clue used. By organizing the existing approaches into a common framework, we hope to illuminate their similarities and differences and establish a firm foundation for future research.

A neural \gls{TSE} system consists of an \gls{NN} that estimates the target speech conditioned on a clue. Fig.~\ref{fig:general_framework} is a schematic diagram of a generic neural \gls{TSE} system that consists of two main modules: a \emph{clue encoder} and a \emph{speech extraction module}, described in more detail below.

\subsection{Clue encoder}
The clue encoder pulls out (from the clue\rev{, $\cluemat$}) information that allows the speech extraction module to identify and extract the target speech in the mixture. We can express the processing as
\begin{align}
    \emb = \clueEnc(\cluemat; \theta^{\text{\rev{C}lue}}),
\end{align}
where $\clueEnc(\cdot; \theta^{\text{\rev{C}lue}})$ represents the clue encoder, which can be an \gls{NN} with learnable parameters $\theta^{\text{\rev{C}lue}}$, and $\emb$ are the clue embeddings. 
Naturally, the specific implementation of the clue encoder and the information carried within $\emb$ largely depend on the type of clues. For example, when the clue is an enrollment utterance, $\emb = \aemb \in \mathbb{R}^{D^{\text{Emb}}}$ will be a speaker embedding vector of dimension $D^{\text{Emb}}$ that represents the voice characteristics of the target speaker. When dealing with visual clues, $\emb = \vemb \in \mathbb{R}^{
 D^{\text{Emb}} \times \N}$ can be a sequence of the embeddings of length $\N$, representing, e.g., the lip movements of the target speaker. Here $\N$ represents the number of time frames of the mixture signal.

Interestingly, the implementation of the speech extraction module does not depend on the type of clues used. To provide a description that is independent of the type of clues, hereafter, we consider that $\emb \in \mathbb{R}^{ D^{\text{Emb}}\times \N}$ consists of a sequence of embedding vectors of dimension $D^{\text{Emb}}$ of length $\N$. Note that we can generate a sequence of embedding vectors for audio clue-based \gls{TSE} systems by repeating the speaker embedding vector for each time frame. 

\subsection{Speech extraction module}

The speech extraction module estimates the target speech from the mixture, given the target speaker embeddings. We can use the same configuration independently of the type of clue. Its process can be decomposed into three main parts: a mixture encoder, a fusion layer, and a target extractor:
\begin{align}
    \Zmix   &= \MixEnc(\mixvec; \theta^{\text{Mix}})\rev{,} \\
    \Ztgt   &= \Fusion(\Zmix,\emb; \theta^{\text{Fusion}})\rev{,} \\
    \estsrcvec[\tgt] & = \Extractor(\Ztgt, \mixvec; \theta^{\Extractor}), 
\end{align}
where $\MixEnc(\cdot; \theta^{\text{Mix}})$, $\Fusion(\cdot; \theta^{\text{Fusion}})$, and $\Extractor(\cdot; \theta^{\Extractor})$ respectively represent the mixture encoder, the fusion layer, and the target extractor with parameters $\theta^{\text{Mix}}$, $\theta^{\text{Fusion}}$, and $\theta^{\Extractor}$. $\Zmix \in \mathbb{R}^{ D^y \times \N}$ and $\Ztgt \in \mathbb{R}^{ D^s \times \N}$ are the internal representations of the mixture before and after conditioning on embedding $\emb$.

The mixture encoder performs the following:
\begin{align}
    \mixfmat &= \FE(\mixvec; \theta^{\text{FE}}), \\
    \Zmix &= \MixNet(\mixfmat; \theta^{\text{MixNet}}),
\end{align}
where $\FE(\cdot)$ and $\MixNet(\cdot)$ respectively represent the feature extraction process and an \gls{NN} with parameters $\theta^{\text{FE}}$ and $\theta^{\text{MixNet}}$. The feature extractor computes the features from the observed mixture signal, $\mixfmat \in \mathbb{R}^{D \times \N}$. These can be such spectral features as magnitude spectrum coefficients derived from the \gls{STFT} of the input mixture \cite{zmolikova2019speakerbeam,wang2018voicefilter,afouras2018conversation,ephrat2018looking}. When using a microphone array, spatial features like \gls{IPD} defined in Eq.~\eqref{eq:ipd} in Section \ref{sec:spatial_tse} can also be appended. Alternatively, the feature extraction process can be implemented by an \gls{NN} such as a 1-D convolutional layer that operates directly on the raw input waveform of the microphone signal \cite{luo2018tasnet,xu2019time}. This enables learning of a feature representation optimized for \gls{TSE} tasks. 

The features are then processed with an \gls{NN}, $\MixNet(\cdot)$, which performs a non-linear transformation and captures the time context, i.e., several past and future frames of the signal. The resulting \rev{representation, $\Zmix$, of the mixture} is (at this point) agnostic of the target.

\begin{table*}[tb]
    \centering
    \caption{Type of fusion layers:  $\W$, $\W_1$, and $\W_2$ are linear transformations for mapping the dimension of \rev{the clue} embeddings\rev{, $D^{\text{Emb}}$,} to the dimension \rev{ of $\Zmix$, $D^Z$}. $\odot$ represents the element-wise Hadamard multiplication operation of matrices. $\mathbf{e}_{i}$ is a vector containing the elements of the $i$-th row of $\emb$ and $\Diag(\cdot)$ is an operator that converts a vector into a diagonal matrix. }
    \begin{tabular}{lll}
        \toprule
        Fusion type & Equation & Parameters ($\theta^{\text{Fusion}}$) \\
        \midrule
        Concatenation & $\Ztgt = \left[\Zmix , \emb\right]$ & - \\
        Addition & $\Ztgt = \Zmix + \W \emb$  & $\W \in \mathbb{R}^{D^Z\times D^{\text{Emb}}}$ \\
        Multiplication & $\Ztgt = \Zmix \odot (\W \emb)$ & $\W \in \mathbb{R}^{D^Z\times D^{\text{Emb}}}$ \\
        \gls{FiLM} & $\Ztgt = \Zmix \odot (\W_1 \emb) + \W_2 \emb,$ & $\W_1 \in \mathbb{R}^{D^Z\times D^{\text{Emb}}}$, $\W_2 \in \mathbb{R}^{D^Z\times D^{\text{Emb}}}$ \\
        Factorized layer & $\Ztgt = \sum_{i=1}^{D^{\text{Emb}}}  \W_i \Zmix \Diag(\mathbf{e}_{i}),$ &  $\W_i \in \mathbb{R}^{D^Z\times D^Z} $\\
        \bottomrule
    \end{tabular}
    \label{tab:fusion_layers}
\end{table*}

The fusion layer, sometimes denoted as an adaptation layer, is a key component of a \gls{TSE} system and allows conditioning of the process on the clue.
It combines \rev{$\Zmix$}  with \rev{the clue} embeddings\rev{, }$\emb$. Conditioning an \gls{NN} on auxiliary information is a general problem that has been studied for multi-modal processing or the speaker adaptation of \gls{ASR} systems. \gls{TSE} systems have borrowed fusion layers from these fields. Table \ref{tab:fusion_layers} lists several options for the fusion layer. Some widely used fusion layers include: (1)~the concatenation of $\Zmix$ with \rev{the clue} embeddings $\emb$\cite{ephrat2018looking,afouras2018conversation}; (2)~addition\footnote{Concatenation is similar to addition if a linear transformation follows it.} after transforming the embeddings with linear transformation $\W$ to match the dimension of $\Zmix$; (3)~multiplication \cite{zmolikova2019speakerbeam}; (4)~a combination of addition and multiplication denoted as \gls{FiLM}; (5)~a factorized layer \cite{zmolikova2017speaker,zmolikova2019speakerbeam}, i.e., the combination of different transformations of the mixture representation weighted by the \rev{clue} embedding values. Other alternatives have also been proposed, including attention-based fusion \cite{xiao2019single}. Note that the fusion operations described here assume just one clue. It is also possible to use multiple clues, as discussed in Section \ref{ssec:mulitmodal}. \rev{Some works also employ the fusion repeatedly at multiple positions in the model \cite{ge2020spex}.}

The last part of the speech extraction module is the target extractor, which estimates the target signal. We explain below the time-frequency masking-based extractor, which has been widely used \cite{zmolikova2017learning,ephrat2018looking,afouras2018conversation,gu2019neural}. Recent approaches also perform a similar masking operation in the learned feature domain\cite{luo2018tasnet,xu2019time}.

\begin{figure*}[tb]
    \centering
    \includegraphics[width=0.8\linewidth]{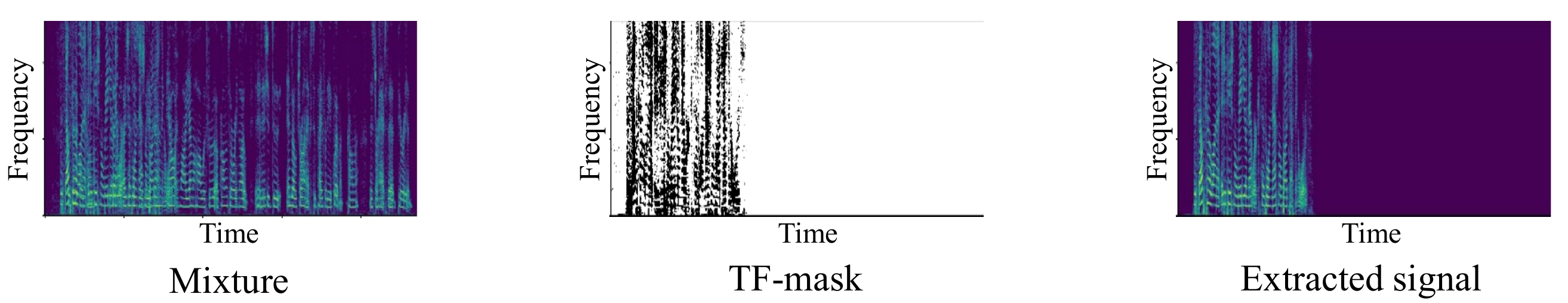}
    \caption{Example of time-frequency mask for speech extraction: Time-frequency mask shows spectrogram regions where target source is dominant. By applying this mask to the mixture, we obtain an extracted speech signal that estimates the target speech.}
    \label{fig:tf_masking}
\end{figure*}

The time-frequency masking approach was inspired by early \gls{BSS} studies that relied on the sparseness assumption of speech signals, an idea based on the observation that the energy of a speech signal is concentrated in a few time-frequency bins of a speech spectrum. Accordingly, the speech signals of different speakers rarely overlap in the time-frequency domain in a speech mixture. We can thus extract the target speech by applying a time-frequency mask on the observed speech mixture, where the mask indicates the time-frequency bins where the target speech is dominant over other signals. 
Fig.~\ref{fig:tf_masking} shows an example of an ideal binary mask for extracting a target speech in a mixture of two speakers. Such an ideal binary mask assumes that all the energy in each TF bin belongs to one speaker. In recent mask-based approaches that use real-valued (or complex) masks, this assumption or observation is not needed.

The processing of the masking-based extractor can be summarized \rev{as} 
\begin{align}
    \maskmt & = \MaskNet(\Ztgt; \theta^{\text{Mask}}), \label{eq:mask_pred}\\   
    \estsrcmt[\tgt] &= \maskmt \odot \mixfmat, \label{eq:masking}\\
    \estsrcvec[\tgt] &= \Reconstruct(\estsrcmt[\tgt]; \theta^{\text{Reconst}}),
\end{align}
where $\MaskNet(\cdot)$ is an \gls{NN} that estimates the time-frequency mask for the target speech, $\maskmt \in \mathbb{R}^{D \times \N}$, $\theta^{\text{Mask}}$ are the network parameters, and $\odot$ denotes the element-wise Hadamard multiplication. $\mixfmat$ and $\estsrcmt[\tgt]$ are the mixture and the estimated target speech signals in the feature domain. Eq.~\eqref{eq:masking} shows the actual extraction process. $\Reconstruct(\cdot)$ is an operation to reconstruct the time-domain signal by performing the inverse operation of the feature extraction of the mixture encoder, i.e., either \gls{iSTFT} or \rev{a} transpose convolution if using a learnable feature extraction. In the latter case, the reconstruction layer has learnable parameters, $\theta^{\text{Reconst}}$.

There are other possibilities to perform the extraction process. For example, we can modify the $\MaskNet(\cdot)$ \gls{NN} to directly infer the target speech signal in the feature domain. Alternatively, as discussed in Section \ref{sec:tse_beamformer}, we can replace the mask-based extraction process with beamforming when a microphone array is available. 

\subsection{Integration with microphone array processing}
\label{sec:tse_beamformer}
If we have access to a microphone array to record the speech mixture, we can exploit the spatial information to extract the target speech. One approach is to use spatial clues to identify the speaker in the mixture by informing the system about the target speaker’s direction, as discussed in Section \ref{sec:spatial_tse}. 
Another approach combines \gls{TSE} with beamforming and uses the latter to perform the extraction process instead of Eq.~\eqref{eq:masking}. For example, we can use the output of a \gls{TSE} system to estimate the spatial statistics needed to compute the coefficients of \rev{a} beamformer steering in the direction of the target speaker. This approach can also be used with audio or visual clue-based \gls{TSE} systems and requires no explicit use of spatial clues to identify the target speaker in the mixture.

We briefly review the mask-based beamforming approach, which was introduced initially for noise reduction and \gls{BSS} \rev{\cite{heymann2016neural,erdogan2016improved}}.
A beamformer performs the linear spatial filtering of the observed microphone signals:
\begin{align}
    \estfnfm =& \mathbf{\BF}^{\mathsf{H}}[\f] \mixfnfmmat,
    \label{eq:bf}
\end{align}
where $\estfnfm \in \mathbb{C}$ is the \gls{STFT} coefficient of the estimated target signal at time frame $\n$ and frequency bin $\f$, $\mathbf{\BF}[\f] \in \mathbb{C}^M$ is a vector of the beamformer coefficients,  $\mixfnfmmat = \left[\mixf^1[\n,\f], \ldots,  \mixf^M[\n,\f] \right]\rev{^T} \in \mathbb{C}^M$ is a vector of the \gls{STFT} coefficients of the microphone signals, $M$ is the number of microphones, and $^{\mathsf{H}}$ is the conjugate transpose.
We can derive the beamformer coefficients from the spatial correlation matrices of the target speech and the interference. These correlation matrices can be computed from the observed signal and the time-frequency mask estimated by the \gls{TSE} system\cite{zmolikova2017speaker}.

This way of combining a \gls{TSE} system with beamforming replaces the time-frequency masking operation of Eq.~\eqref{eq:masking} with the spatial linear filtering operation of Eq.~\eqref{eq:bf}. It allows distortionless extraction, which is often advantageous when using \gls{TSE} as a front-end for \gls{ASR} \cite{zmolikova2019speakerbeam}.

\subsection{Training a TSE system}
\label{sec:ssec_training}
Before using a \gls{TSE} model, we first need to learn its parameters: $\theta^{\text{TSE}} =\{\theta^{\text{Mix}}, \theta^{\text{Clue}}, \theta^{\text{Fusion}},\theta^{\text{TgtExtractor}}\}$. Most existing studies use fully supervised training, which requires a large amount of training data consisting of the triplets of speech mixture $\mixvec$, target speech signal $ \srcvec[\tgt]$, and corresponding clue $\cluemat$ to learn parameters $\theta^{\text{TSE}}$. Since this requires access to a clean target speech signal, such training data are usually simulated by artificially mixing clean speech signals and noise following the signal model of Eq.~\eqref{eq:mixt_1ch_time}. 

\begin{figure*}[tb]
    \centering
    \includegraphics[width=0.7\linewidth]{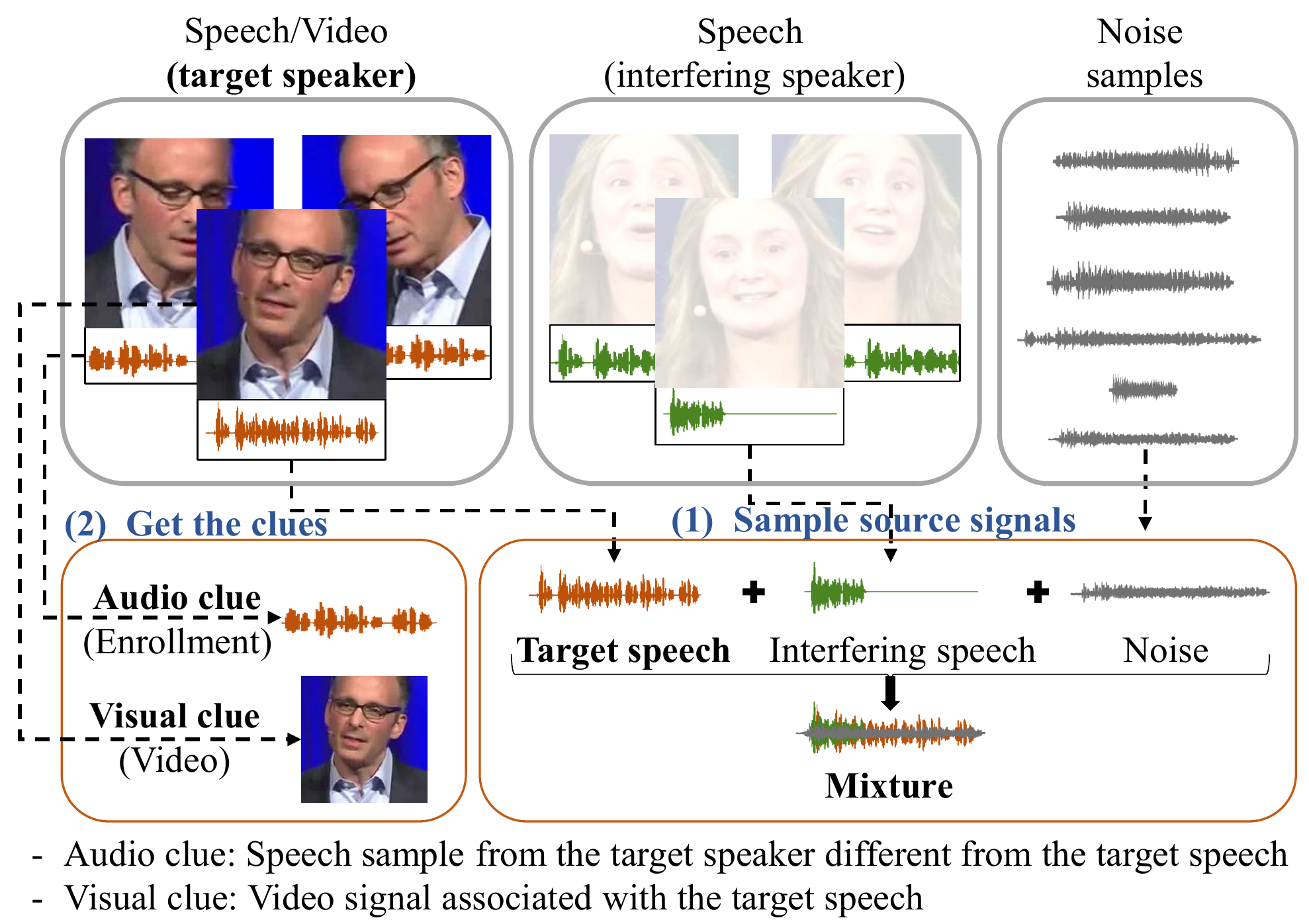}
    \caption{Example of generating simulation data for training or testing: This example assumes videos are available so that audio and visual clues can be generated. No video is needed for audio clue-based \gls{TSE}. For visual clue-based \gls{TSE}, we do not necessarily need multiple videos from \rev{the} same speaker.}
    \label{fig:data_generation}
\end{figure*}

Figure~\ref{fig:data_generation} illustrates the data generation process using a multi-speaker audio-visual speech corpus containing multiple videos for each speaker. First, we generate a mixture using randomly selected speech signals from the target speaker, the interference speaker, and the background noise. We obtain an audio clue by selecting another speech signal from the target speaker as well as a visual clue from the video signal associated with the target speech.   

The training of a neural \gls{TSE} framework follows the training scheme of \glspl{NN} with error back-propagation. The parameters are estimated by minimizing a training loss function:

\begin{align}
    \theta^{\text{TSE}} = \argmin_\theta \mathcal{L}\left(\srcvec[\tgt], \estsrcvec[s] \right),
\end{align}
where $\mathcal{L}(\cdot)$ is a training loss, which measures how close estimated target speech $\estsrcvec[s]\ \rev{ =\TSE\left(\mixvec, \cluemat; \theta\right)}$ is to the target source signal $\srcvec[\tgt]$. We can use a similar loss as that employed for training noise reduction or \gls{BSS} systems \cite{wang2018supervised,luo2018tasnet}. 

Several variants of the losses operating on different domains exist, such as the cross-entropy between the oracle and the estimated time-frequency masks and the \gls{MSE} loss between the magnitude spectra of the source and the estimated target speech. Recently, a negative \gls{SNR} measured in the time-domain has been widely used \cite{luo2018tasnet,xu2019time,michelsanti2021overview}:
\begin{align}
    \mathcal{L}^{\text{SNR}} (\srcvec[\tgt], \estsrcvec[s] )& =  - 10 \log_{10} \left( \frac{\| \srcvec[\tgt] \|^{2}}{\| \srcvec[\tgt] - \estsrcvec[s] \|^{2}} \right). \label{eq:sdr_loss} 
\end{align}
The \gls{SNR} loss is computed directly in the time-domain, which forces the \gls{TSE} system to learn to correctly estimate the magnitude and the phase of the target speech signal. This loss has improved extraction performance \cite{xu2019time}. Many works also employ versions of the loss which are invariant to arbitrary scaling, i.e., \gls{SI-SNR}\rev{\cite{luo2018tasnet}} or linear filtering of the estimated signal, often called\gls{SDR} \rev{\cite{vincent2006performance}}.
Besides training losses operating on the signal or mask levels, it is also possible to train a \gls{TSE} system end-to-end with a loss defined on the output of an \gls{ASR} system\cite{delcroix2019end}. Such a loss can be particularly effective when targeting \gls{ASR} applications, as discussed in Section \ref{sec:extension}.

The clue encoder can be an \gls{NN} trained jointly with a speech extraction module \cite{zmolikova2019speakerbeam} or pre-trained on a different task, such as speaker identification for audio clue-based \gls{TSE} \cite{wang2018voicefilter} or lip-reading for visual clue-based \gls{TSE} \cite{afouras2018conversation}. Using a pre-trained clue encoder enables the leveraging of large amounts of data to learn robust and highly discriminative embeddings. On the other hand, jointly optimizing the clue encoder allows learning embeddings to be optimized directly for \gls{TSE}. These two trends can also be combined by fine-tuning the pre-trained encoder or using multi-task training schemes, which add \rev{a} loss to the output of the clue embeddings\cite{mun2020sound}.

\subsection{Considerations when designing a TSE system}
We conclude this section with some considerations about the different options for designing a \gls{TSE} system. In the above description, we intentionally ignored the details of the \gls{NN} architecture used in the speech extraction module, such as the type of layers. Indeed, novel architectures have been and will probably continue to be proposed regularly, leading to gradual performance improvement. For concrete examples, we refer to some public implementations of \gls{TSE} frameworks presented in Section \ref{sec:resources}.

Most \gls{TSE} approaches borrow a network configuration from architectures proven effective for \gls{BSS} or noise reduction.
One important aspect is that an \gls{NN} must be able to see enough context in the mixture to identify the target speaker. This has been achieved using such \gls{RNN}-based architectures as a stack of \gls{BLSTM} layers \cite{zmolikova2019speakerbeam}, \gls{CNN}-based architectures with a stack of convolutional layers that gradually increases the receptive field over the time axis to cover a large context\cite{afouras2018conversation,xu2019time} or attention-based architectures\cite{li21c_interspeech}. 

The networks in the mixture encoder and the extraction process generally use a similar architecture. The best performance was reported when using a shallow mixture encoder (typically a single layer/block) and a much deeper extraction network, i.e., where a fusion layer is placed on the lower part of the extraction module. Furthermore, we found in our experiments that the multiplication or \gls{FiLM} layers usually perform well. However, the impact of the choice of the fusion layer seems rather insignificant.

For the feature extraction, early studies used spectral features computed with \gls{STFT}\cite{zmolikova2019speakerbeam,afouras2018conversation,ephrat2018looking}. However, most recent approaches employ a learned feature extraction module following its success for separation \cite{luo2018tasnet,xu2019time}. \rev{This approach allows direct optimization of the features for the given task.} However, the choice of input features may depend on the acoustic conditions, and some have reported superior performance using \gls{STFT} under challenging reverberant conditions\rev{\cite{cord2022monaural} or using handcrafted filterbanks \cite{ditter2020multi}}.

Except for such general considerations, it is difficult to make solid arguments for a specific network configuration since performance may depend on many factors, such as the task, the type of clue, the training data generation, and the network and training hyper-parameters.

\section{Audio-based TSE}
\label{sec:audio_tse}
 
In this section, we explain how the general framework introduced in Section \ref{sec:general_framework} can be applied in the case of audio clues. In particular, we discuss different options to implement the clue encoder, summarize the development of the audio-based \gls{TSE}, and present some representative experimental results. 

\subsection{Audio clue encoder}
\label{ssec:audio_clue_encoder}
\begin{figure*}[tb]
    \centering
    \includegraphics[width=.8\textwidth]{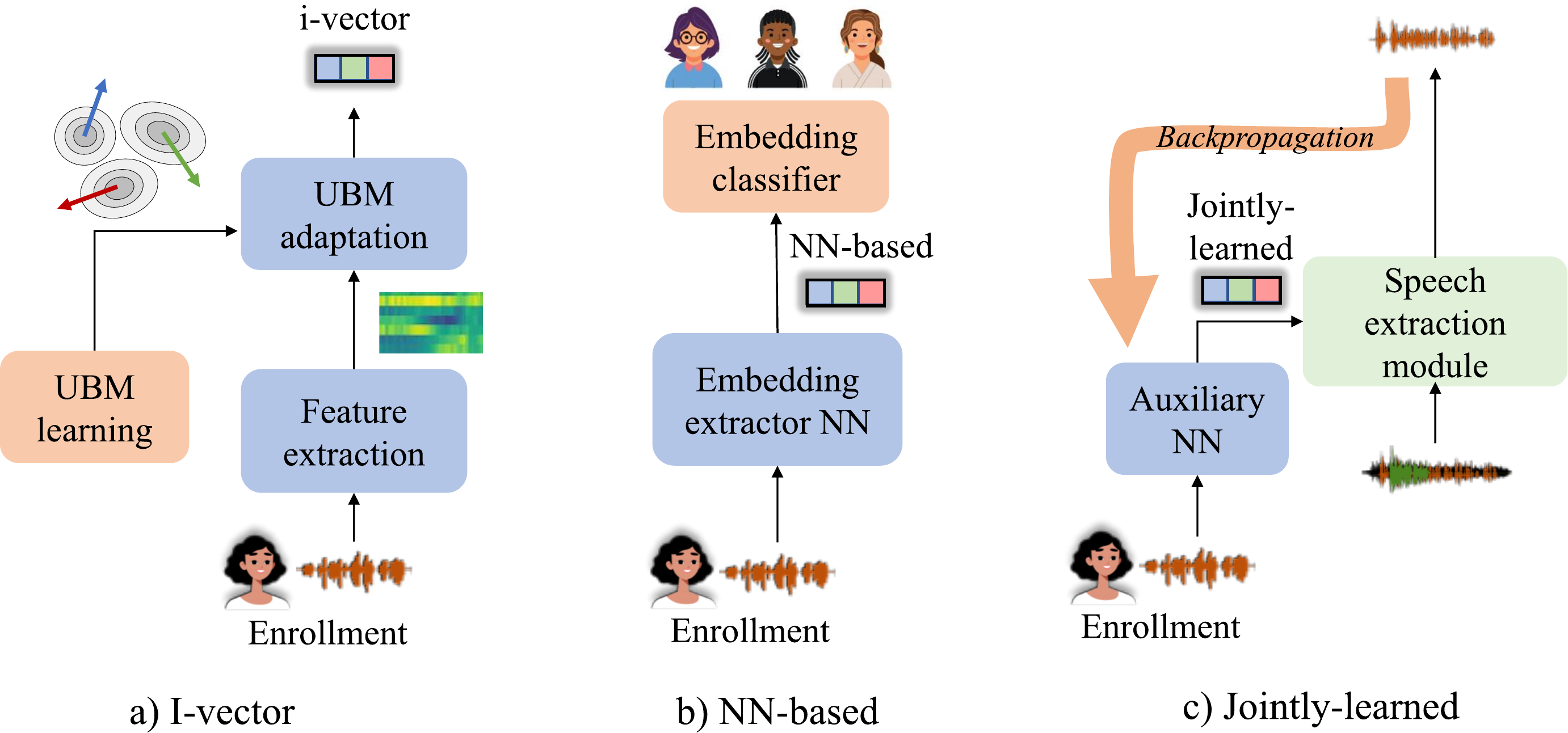}
    \caption{Illustration of i-vector, \gls{NN}-based vector, and jointly-trained embeddings: Orange parts are included only in training stage.}
    \label{fig:audio_clue_encoder}
\end{figure*}
An audio clue is an utterance spoken by the target speaker from which we derive the characteristics of \rev{their} voice, allowing identification in a mixture. This enrollment utterance can be obtained by pre-recording the user of a personal device or with a part of a recording where a wake-up keyword was uttered. The clue encoder is usually used to extract a single vector that summarizes the entire enrollment utterance. 

Since the clue encoder’s goal is to extract information that defines the voice characteristics of the target speaker, embeddings from the speaker verification field are often used, such as i-vectors or \gls{NN}-based embeddings (e.g., d-vectors or x-vectors). Clue encoders trained directly for \gls{TSE} tasks are also used. Fig.~\ref{fig:audio_clue_encoder} describes these three options.

\subsubsection{I-vectors}
From their introduction around 2010, i-vectors \cite{dehak2010front} were the ruling speaker verification paradigm until the rise of \gls{NN} speaker embeddings. The main idea behind i-vectors is modeling the features of an utterance using a \gls{GMM}, whose means are constrained to a subspace and depend on the speaker and the channel effects. The subspace is defined by the \gls{UBM}, i.e., \gls{GMM} trained on a large amount of data from many speakers, and a total variability subspace matrix. The super-vector of the means of utterance \gls{GMM} $\boldsymbol{\mu}$ is decomposed:
\begin{equation}
    \boldsymbol{\mu} = \mathbf{m} + \mathbf{T}\mathbf{w},
\end{equation}
where $\mathbf{m}$ is a super-vector of the means of the \gls{UBM}, $\mathbf{T}$ is a low-rank rectangular matrix representing the bases spanning the subspace, and $\mathbf{w}$ is a random variable with standard normal prior distribution. Since an i-vector is the maximum a posteriori estimate of $\mathbf{w}$, it thus consists of values that enable the adaptation of the parameters of the generic \gls{UBM} speaker model ($\mathbf{m}$) to a specific recording. As a result, it captures the speaker’s voice characteristics in the recording.

An important characteristic of i-vectors is that they capture both the speaker and channel variability. This case may be desired in some \gls{TSE} applications, where we obtain enrollment utterances in identical conditions as the mixed speech. In such a situation, the channel information might also help distinguish the speakers. I-vectors have also been used in several \gls{TSE} works \cite{zmolikova2019speakerbeam}.

\subsubsection{Neural network-based embeddings}
\rev{The state-of-the-art speaker verification systems predominantly use \gls{NN}-based speaker embeddings, which were adopted later for \gls{TSE}.}
The common idea is to train an \gls{NN} for the task of speaker classification. Such an \gls{NN} contains a ``pooling layer'' which converts a sequence of features into one vector. The pooling layer computes the mean and \rev{optionally} the standard deviation of the sequence of features over the time dimension. The pooled vector is then classified into speaker classes or used in other loss functions that encourage speaker discrimination. For \gls{TSE}, the speaker embedding is then the vector of the activation coefficients of one of the last network layers. \rev{The most common of such \gls{NN}-based speaker embeddings are d-vectors and x-vectors \cite{snyder2018xvectors}. Many \gls{TSE} works employ d-vectors \cite{wang2018voicefilter}.}

Since  \glspl{NN} are trained for speaker classification or a related task, embeddings are usually highly speaker-discriminative. Most other sources of variability are discarded, such as the channel or content. Another advantage of this class of embeddings is that they are usually trained on large corpora with many speakers, noises, and other variations, resulting in very robust embedding extractors. Trained models are often publicly available, and the embeddings can be readily used for \gls{TSE} tasks.

\subsubsection{Jointly-learned embeddings}
\label{ssec:jointly_learned}
\gls{NN}-based embeddings, such as x-vectors, are designed and trained for the task of speaker classification. Although this causes them to contain speaker information, it is questionable whether the same representation is optimal for \gls{TSE} tasks. An alternative is to train the neural embedding extractor jointly with a speech extraction module. The resulting embeddings are thus directly optimized for \gls{TSE} tasks.  This approach has been used for \gls{TSE} in several works\cite{zmolikova2019speakerbeam,ge2020spex}. 

The \gls{NN} performing the speaker embedding extraction takes \rev{an} enrollment utterance $\acluemat$ as input and generally contains a pooling layer converting the frame-level features into one vector, similar to the embedding extractors discussed above. This \gls{NN} is trained with the main \gls{NN} using a common objective function. A second objective function can also be used on the embeddings to improve their speaker discriminability \cite{mun2020sound}.

As mentioned above, the advantage of such embeddings is that they are trained directly for \gls{TSE} and thus collect essential information for this task. On the other hand, the pre-trained embedding extractors are often trained on larger corpora and may be more robust. A possible middle ground might take a pre-trained embedding extractor and fine-tune it jointly with the \gls{TSE} task. However, this has, to the best of our knowledge, not been done yet.

\subsection{Existing approaches}
The first neural \gls{TSE} methods were developed around 2017. One of the first published works, SpeakerBeam \cite{zmolikova2019speakerbeam}, explored both the single-channel approach, where the target extractor was implemented by time-frequency masking, and the multi-channel approach using beamforming. This work also compared different variants of fusion layers and clue encoders. This was followed by VoiceFilter \cite{wang2018voicefilter}, which put more emphasis on \gls{ASR} applications using \gls{TSE} as a front-end and also investigated streaming variants with minimal latency. A slightly modified variant of the task was presented in works on speaker inventory \cite{xiao2019single}, where not one but multiple speakers can be enrolled. Such a setting might be suitable for meeting scenarios. Recently, many works, such as SpEx \cite{ge2020spex}, have started to use time-domain approaches, following their success in \gls{BSS}\cite{luo2018tasnet}. 

\subsection{Experiments}
\label{ssec:audio_tse_exp}

\begin{figure}[tb]
    \centering
    \includegraphics[width=0.45\textwidth]{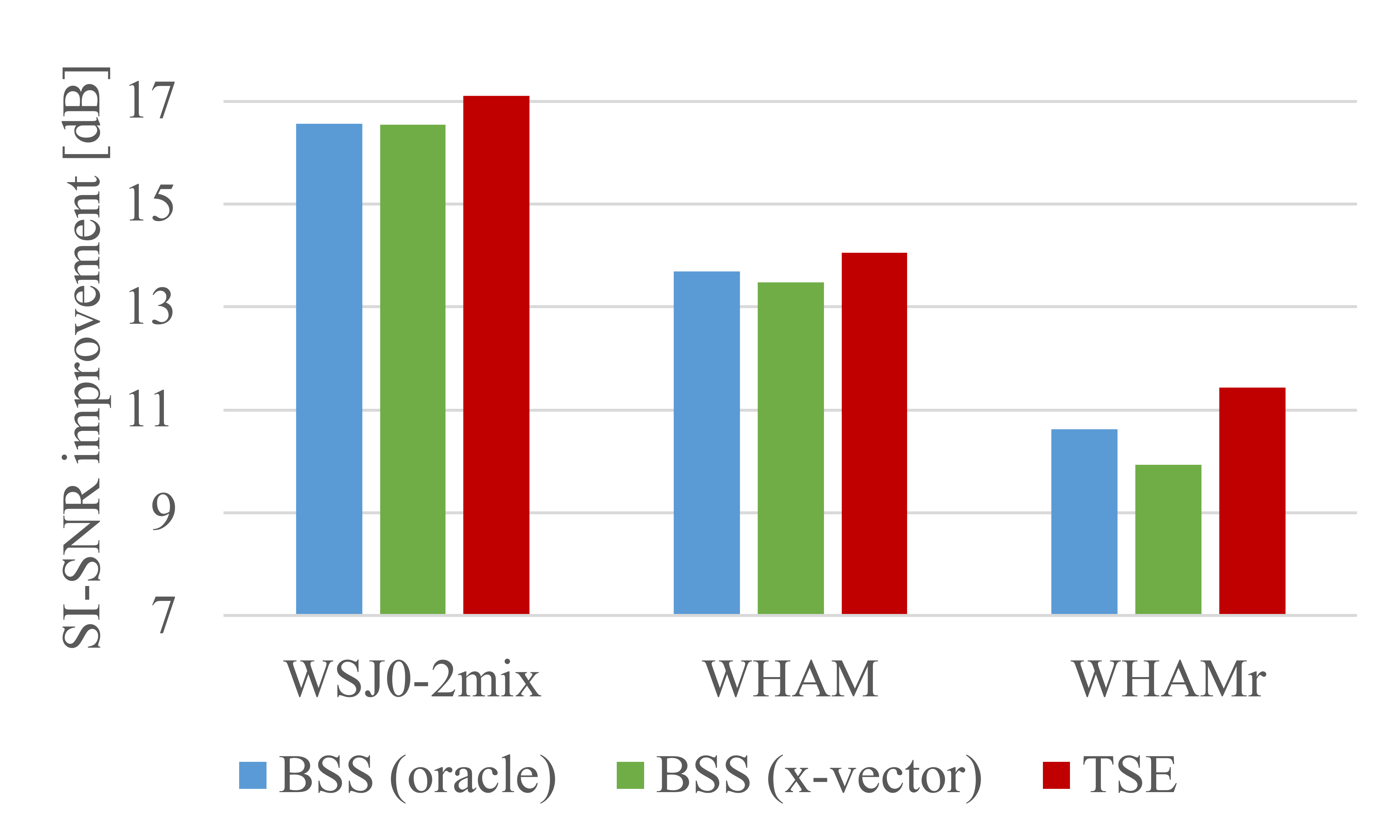}
    \caption{Comparison of \gls{TSE} and cascade \gls{BSS} systems when using an audio clue in terms of \gls{SI-SNR} improvement (higher is better) \rev{\cite{zmolikova2022neural}}.}
    \label{fig:audio_results}
\end{figure}

\begin{figure*}[tb]
    \centering
    \includegraphics[width=0.65\textwidth]{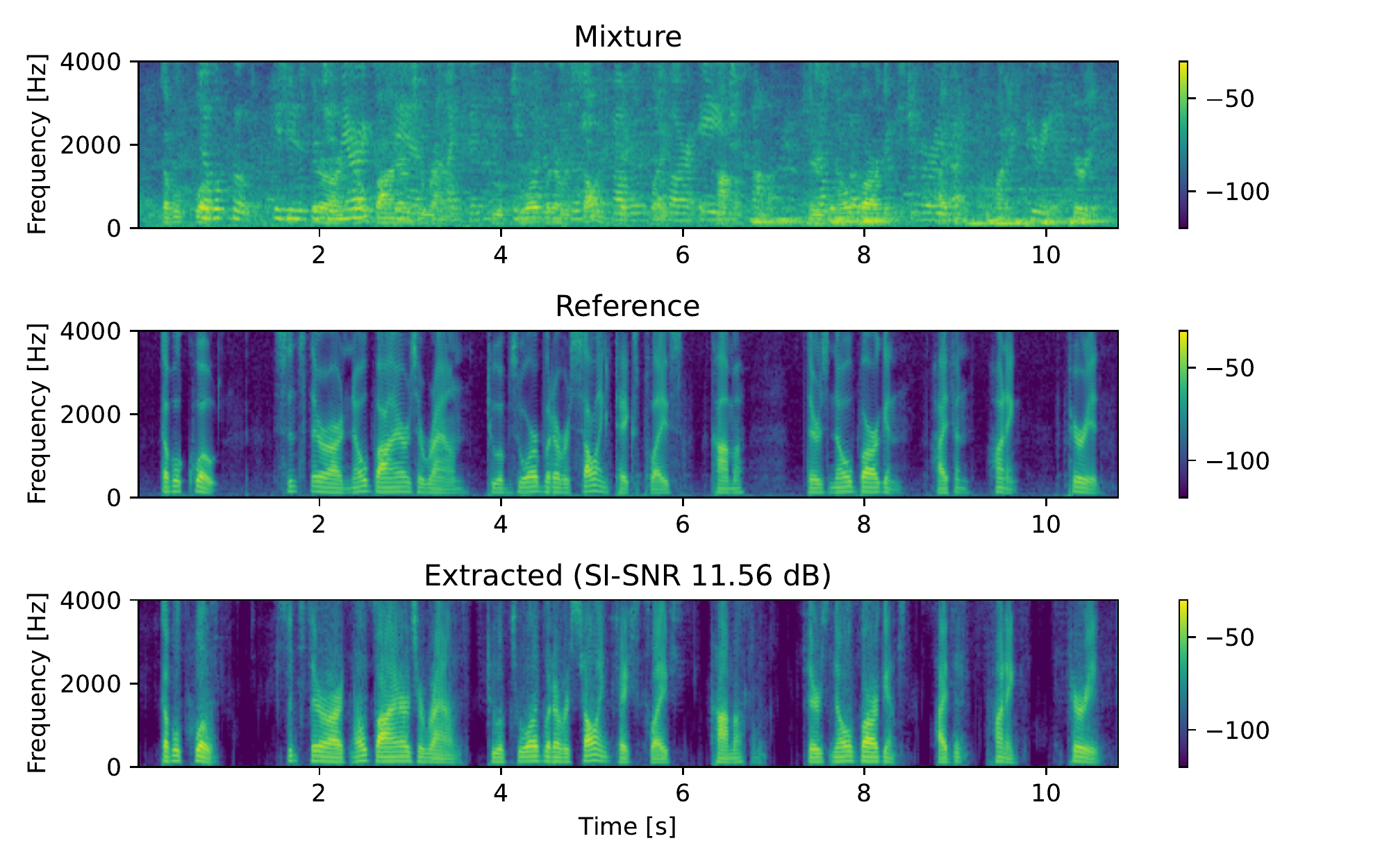}
    \caption{Example of spectrograms of mixed, reference, and extracted speech: Example is taken from WHAM\rev{R!} database.}
    \label{fig:specgrams}
\end{figure*}

An audio clue is a simple way to condition the system for extracting the target speaker. Many works have shown that the speaker information extracted from audio clues is sufficient for satisfactory performance. Demonstrations of many works are available online\footnote{Demonstrations of audio clues approaches: VoiceFilter \cite{wang2018voicefilter} \url{https://google.github.io/speaker-id/publications/VoiceFilter/}, SpeakerBeam \cite{zmolikova2019speakerbeam} \url{https://www.youtube.com/watch?v=7FSHgKip6vI}.}. We present here some results to demonstrate the potential of audio clue-based approaches. The experiments were done with time-domain SpeakerBeam\footnote{\url{https://github.com/butspeechfit/speakerbeam}}, which uses a convolutional architecture, a multiplicative fusion layer, and a jointly-learned clue encoder.

The experiments were done on three different datasets (WSJ0-2mix, WHAM\rev{!}, and WHAM\rev{R!}) to show the performance in different conditions (clean, noisy, and reverberant, respectively). We describe these datasets in more detail in Section~\ref{sec:resources}. All the experiments were evaluated with the \gls{SI-SNR} metric and measured the improvements over the \gls{SI-SNR} of the observed mixture. \rev{More details about the experiments can be found in \cite{zmolikova2022neural}.}

Figure \ref{fig:audio_results} compares the \gls{TSE} results with a cascade system, first doing \gls{BSS} and then independent speaker identification. Speaker identification is done either in an oracle way (selecting the output closest to the reference) or with x-vectors (extracting the x-vectors from all the outputs and the enrollment utterances and selecting the output with the smallest cosine distance as the target). \rev{The \gls{BSS} system uses the same convolutional architecture as \gls{TSE}, differing only in that it does not have a clue encoder and the output layer is twice larger as it outputs two separated speech signals.} The direct \gls{TSE} scheme outperformed the cascade system, especially in more difficult conditions such as WHAM\rev{R!}. This difference reflects a couple of causes: 1) the \gls{TSE} model is directly optimized for the \gls{TSE} task and does not spend any capacity on extracting other speakers or 2) the \gls{TSE} model has additional speaker information.

Figure \ref{fig:specgrams} shows an example of spectrograms obtained using \gls{TSE} on a recording of two speakers from the WHAM\rev{R!} database, including noise and reverberation. \gls{TSE} correctly identifies the target speaker and removes all the interference, including the second speaker, noise, and reverberation.

\subsection{Limitations and outlook}
Using \gls{TSE} systems conditioned on audio clues is particularly practical due to the simplicity of obtaining the clues, i.e., no additional hardware is needed, such as cameras or multiple microphones. \rev{Considering} the good performance demonstrated in the literature, these systems are widely applicable. Nowadays, the methods are rapidly evolving and achieving increasingly higher accuracy.

The main challenge in audio-clue-based systems is correct identification of the target speaker. The speech signal of the same speaker might have highly different characteristics in different conditions due to such factors as emotional state, channel effects, or the Lombard effect. \gls{TSE} systems must be robust enough to such intra-speaker variability. On the other hand, different speakers might have very similar voices, leading to erroneous identification if the \gls{TSE} system lacks sufficient accuracy.

Resolving both issues requires precise speaker modeling. In this regard, the \gls{TSE} methods may draw inspiration from the latest advances in the speaker verification field, including advanced model architectures, realistic datasets with a huge number of speakers for training, or using pre-trained features from self-supervised models.

\section{Visual/Multi-modal clue-based TSE}
\label{sec:visual_tse}

Visual clue-based \gls{TSE} assumes that a video camera captures the face of the target speaker who is talking in the mixture \cite{afouras2018conversation,ephrat2018looking}. Using visual clues is motivated by psycho-acoustic studies (see the references in a previous work \cite{michelsanti2021overview}) that revealed that humans look at lip movements to understand speech better. Similarly, the visual clues of \gls{TSE} systems derive hints about the state of the target speech from the lip movements, such as whether the target speaker is speaking or silent or more refined information about the phoneme being uttered.  

A visual clue, which presents different characteristics than audio clues because it captures information from another modality, is time-synchronized with the target speech in the mixture without being corrupted by the interference speakers. Therefore, a visual clue-based \gls{TSE} can better handle mixtures of speakers with similar voices, such as same-gender mixtures, than audio clue-based systems because the extraction process is not based on the speaker’s voice characteristics\footnote{Some works can even perform extraction from a mixture of the same speaker’s speech\cite{ephrat2018looking}.}. Another potential advantage is that the users may not need to pre-enroll their voice. Video signals are also readily available for many applications such as video-conferencing.

\begin{figure*}[tb]
    \centering
    \includegraphics[width=0.7\linewidth]{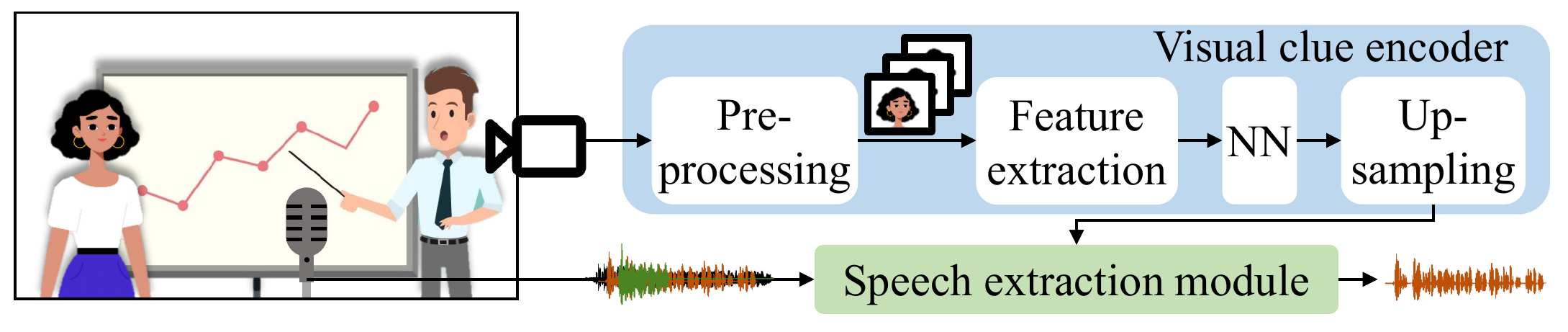}
    \caption{Visual clue-based \gls{TSE} system.}
    \label{fig:visual_tse}
\end{figure*}

Figure \ref{fig:visual_tse} shows a diagram of a visual \gls{TSE} system that follows the same structure as the general \gls{TSE} framework introduced in Section \ref{sec:general_framework}. Only the visual clue encoder part is specific to the task. We describe it in more detail below and then introduce a multi-modal clue extension. We conclude this section with some experimental results and discussions.

\subsection{Visual clue encoder}
\label{ssec:visual_clue_encoder}
The visual clue encoder computes from the video signal a representation that allows the speech extraction module to identify and extract the target speech in the mixture. This processing involves the steps described below:
\begin{align}
\vemb = \upsample(\NN(\VFE(\vcluemat), \theta^{\text{v-clue}})),
\end{align}
where $\vemb \in \mathbb{R}^{D^{\text{Emb}} \times \N }$ represents the sequence of the visual embedding vectors, $\vcluemat$ is the video signal obtained after pre-processing, $\VFE(\cdot)$ is the visual feature extraction module, $\NN(\cdot, \theta^{\text{v-clue}})$ is an \gls{NN} with parameters $\theta^{\text{v-clue}}$, and $\upsample(\cdot)$ represents the up-sampling operation. The latter up-sampling step is required because the sampling rates of the audio and video devices are usually different. Up-sampling matches the number of frames of the mixture and visual clue encoders. 

\subsubsection{Pre-processing}
First, the video signal captured by the camera requires pre-processing to isolate the face of the target speaker. Depending on the application, this may require detecting and tracking the target speaker’s face and cropping the video. These pre-processing steps can be performed using previously well-established video processing algorithms\cite{michelsanti2021overview}. 
\subsubsection{Visual feature extraction}
Similar to an audio-clue-based \gls{TSE}, the visual clue encoder can directly extract embeddings from raw video data or visual features. With the first option, the raw video is processed with a \gls{CNN} whose parameters are jointly-learned with the speech extraction module to enable direct optimization of the features for the extraction task without any loss of information. However, since the video signals are high-dimensional data, achieving joint optimization can be complex. This approach has been used successfully with speaker-close conditions \cite{gabbay2018visual}. Extending it to speaker-open conditions might require a considerable amount of data or careful design of the training loss using, e.g., multi-task training to help the visual encoder capture relevant information. 

Most visual \gls{TSE} works use instead a visual feature extractor pre-trained on another task to reduce the dimensionality of the data. Such feature extractors can leverage a large amount of image or video data (that do not need to be speech mixtures) to learn representation robust to variations, such as resolution, luminosity, or head orientation. 
The first option is to use facial landmark points as features. Facial landmarks are the key points on a face that indicate the mouth, eyes, or nose positions and offer a very low-dimension representation of a face, which is interpretable. Moreover, face landmarks can be easily computed with efficient off-the-shelf algorithms \cite{morrone2019face}.  

The other option is to use neural embeddings derived from an image/video processing \gls{NN} trained on a different task, which was proposed in three concurrent works \cite{ephrat2018looking,afouras2018conversation,owens2018audio}. Ephrat et al. \cite{ephrat2018looking} used visual embeddings obtained from an intermediate layer of a face recognition system called FaceNet. This face recognition system is trained so that embeddings derived from photographs of the same person are close and embeddings from different persons are far from each other. It thus requires only a corpus of still images with person identity labels for training the system. However, the embeddings do not capture the lip movement dynamics and are not explicitly related to the acoustic content. 

Alternatively, Afouras et al. \cite{afouras2018conversation} proposed using embeddings obtained from a network trained to perform lip-reading, i.e., where a network is trained to estimate the phoneme or word uttered from the video of the speaker’s lips. The resulting embeddings are thus directly related to the acoustic content. However, the training requires video with the associated phoneme or word transcriptions, which are more demanding and costly to obtain.

The third option introduced by Owens et al. \cite{owens2018audio} exploits embeddings derived from an \gls{NN} trained to predict whether the audio and visual tracks of a video are synchronized. This approach enables self-supervised training, where the training data are simply created by randomly shifting the audio track by a few seconds. The embeddings capture information on the association between the lip motions and the timing of the sounds in the audio. 
All three options \cite{ephrat2018looking,afouras2018conversation,owens2018audio} can successfully perform a visual \gls{TSE}.

\subsubsection{Transformation and up-sampling}
Except with joint-training approaches, the visual features are (pre-)trained on different tasks and thus do not provide a representation optimal for \gls{TSE}. Besides, since some of the visual features are extracted from the individual frames of a video, the dynamics of lip movements are not captured.
Therefore, the visual features are further transformed with an \gls{NN}, which is jointly trained with the speech extraction module. The \gls{NN}, which allows learning a representation optimal for \gls{TSE}, can be implemented with \gls{LSTM} or convolutional layers across the time dimension to model the time series of the visual features, enabling the lip movement dynamics to be captured.
Finally, the visual embeddings are up-sampled to match the sampling rate of audio features $\Zmix$. 

\subsection{Audio-visual clue-based TSE}
\label{ssec:mulitmodal}
Audio and visual clue-based \gls{TSE} systems have complementary properties. An audio clue-based \gls{TSE} is not affected by speaker movements and visual occlusions. In contrast, a visual clue-based \gls{TSE} is less affected by the voice characteristics of the speakers in the mixture. By combining these approaches, we can build \gls{TSE} systems that exploit the strengths of both clues for improving the robustness to various conditions\cite{ochiai2019multimodal,gu2020multi}. 

\begin{figure}[tb]
    \centering
    \includegraphics[width=1\linewidth]{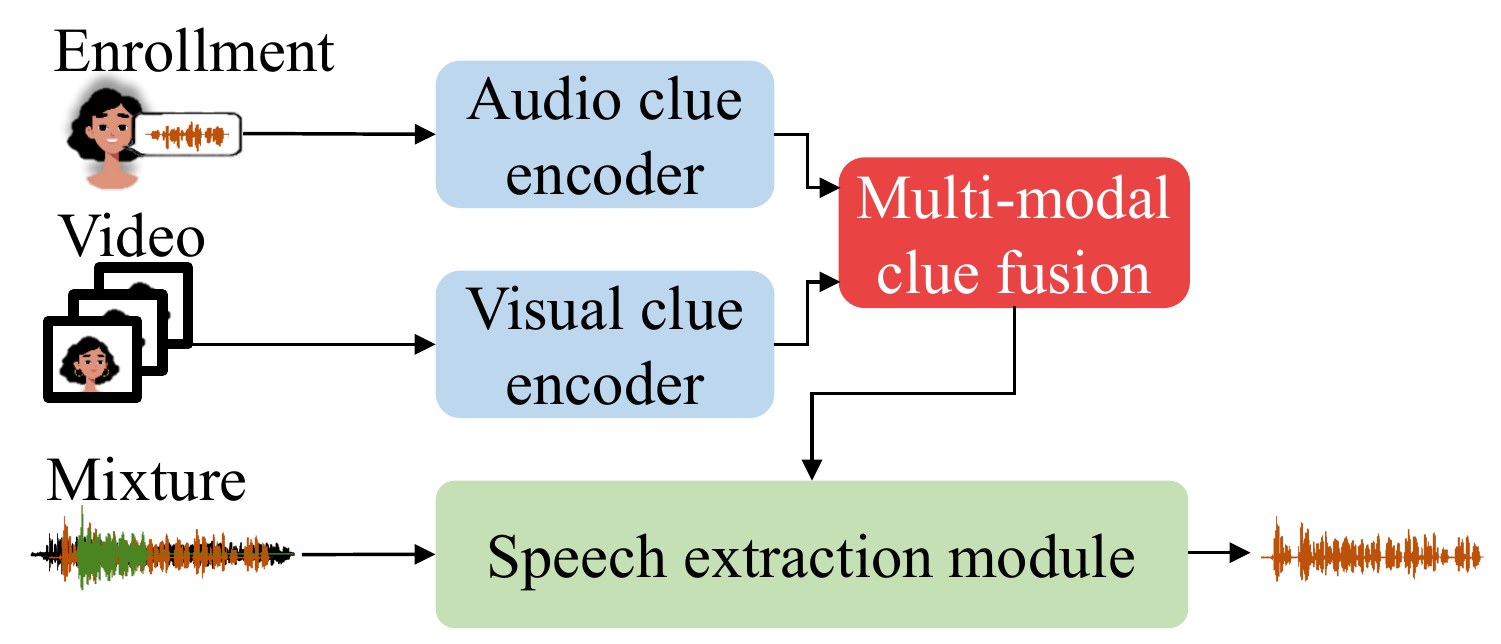}
    \caption{Audio-visual clue-based \gls{TSE} system }
    \label{fig:av_tse}
\end{figure}

Figure \ref{fig:av_tse} shows a diagram of an audio-visual \gls{TSE} system, which assumes access to the pre-recorded enrollment of the target speaker to provide an audio clue and a video camera for a visual clue. The system uses the audio and visual clue encoders described in Sections \ref{ssec:audio_clue_encoder} and \ref{ssec:visual_clue_encoder} and combines these clues into an audio-visual embedding, which is given to the speech extraction module. Audio-visual embeddings can be simply the concatenation \cite{afouras2019my} \rev{or} the summation of the audio and visual embeddings, or obtained as a weighted sum \cite{ochiai2019multimodal,sato2021multimodal}, where the weights can vary depending on the reliability of each clue. The weighted sum approach can be implemented with an attention layer widely used in machine learning, which enables dynamic weighting of the contribution of each clue.
 
\subsection{Experimental results and discussion}
Several visual \gls{TSE} systems have been proposed, which differ mostly by the type of visual features used and the network configuration. These systems have demonstrated astonishing results, which can be attested by the demonstrations available online\footnote{Demo samples for several approaches are available, e.g., for \cite{owens2018audio}: \url{https://andrewowens.com/multisensory}, for \cite{ephrat2018looking}:  \url{https://looking-to-listen.github.io}, for \cite{afouras2018conversation}: \url{https://www.robots.ox.ac.uk/~vgg/demo/theconversation}, and for \cite{sato2021multimodal}: \url{http://www.kecl.ntt.co.jp/icl/signal/member/demo/audio_visual_speakerbeam.html}}. Here we briefly describe experiments using the audio, visual, and audio-visual time-domain SpeakerBeam systems\cite{sato2021multimodal}, which use a similar configuration as the system in Section \ref{ssec:audio_tse_exp}. The speech extraction module employs a stack of time-convolutional blocks and a multiplicative fusion layer. The audio clue encoder consists of the jointly-learned embeddings described in Section \ref{ssec:jointly_learned}. The visual clue encoder uses visual features derived from face recognition like a previous work \cite{ephrat2018looking}. The audio-visual system combines the visual and audio clues with an attention layer\cite{sato2021multimodal}.

The experiments used mixtures of utterances from the LRS3-TED corpus\footnote{\url{https://www.robots.ox.ac.uk/~vgg/data/lip_reading/lrs3.html}}, which consists of single speaker utterances with associated videos. We analyzed the behavior under various conditions by looking at results from same and different gender mixtures and two examples of clue corruptions (enrollment corrupted with white noise at \gls{SNR} of 0~dB and video with a mask on the speaker’s mouth). The details of the experimental setup are available in \cite{sato2021multimodal}. 

\begin{figure*}[tb]
    \centering
    \includegraphics[width=0.75\linewidth]{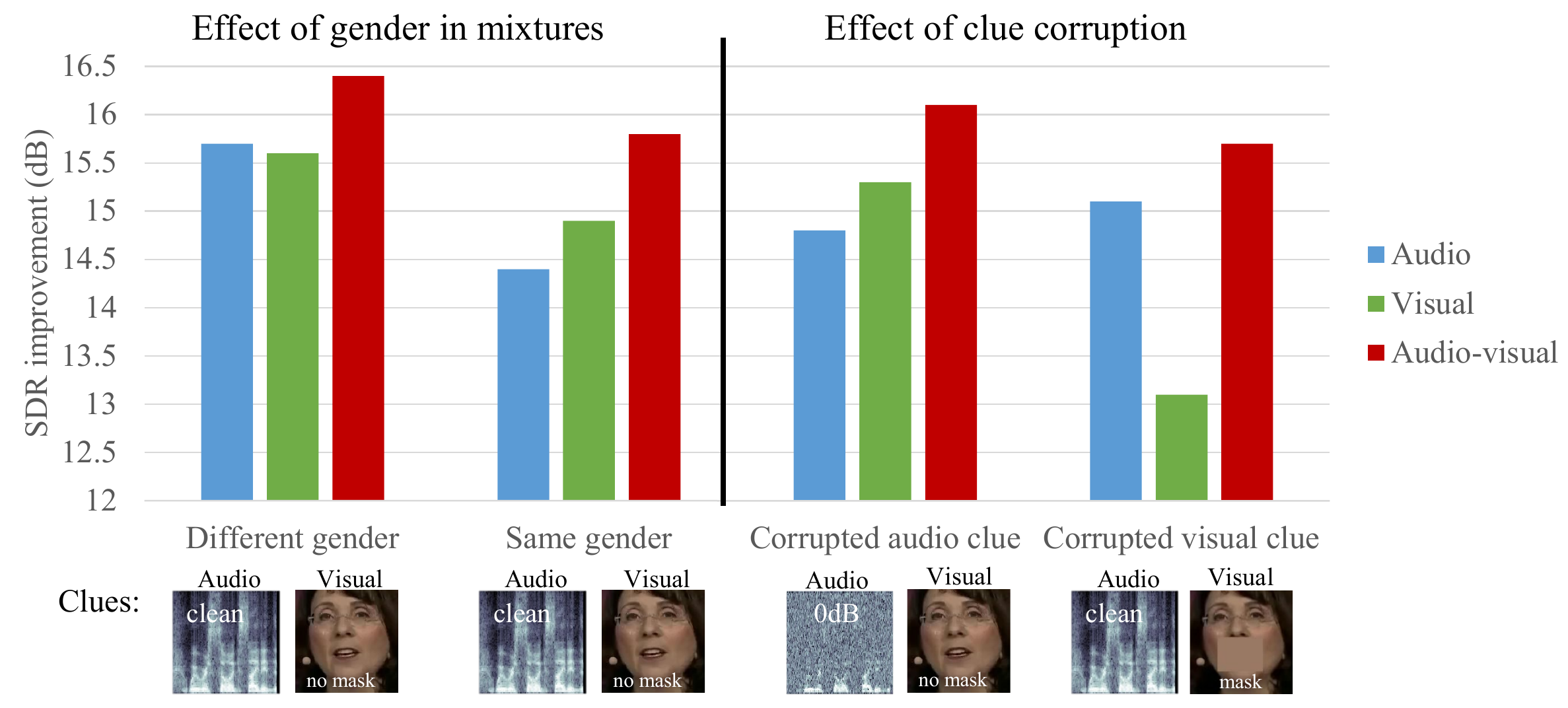}
    \caption{SDR Improvement of \gls{TSE} with audio, visual, and audio-visual clues for mixtures of same/different gender and for corruptions of audio and visual clues: Audio clues were corrupted by adding white noise at \gls{SNR} of 0 dB to enrollment utterance. Video clues were corrupted by masking mouth region in video. }
    \label{fig:av_results}
\end{figure*}
Figure \ref{fig:av_results} compares the extraction performance measured in terms of \gls{SDR} improvement for audio, visual, and audio-visual \gls{TSE} under various mixture and clue conditions. We confirmed that a visual clue-based \gls{TSE} is less sensitive to the characteristics of the speakers in the mixture since the performance gap between different- and same-gender mixtures is smaller than with an audio clue-based \gls{TSE}. When using a single clue, performance can be degraded when this clue is corrupted. However, the audio-visual system that exploits both clues can achieve superior extraction performance and is more robust to clue corruption.

\subsection{Discussions and outlook}
Visual clue-based \gls{TSE} approaches offer an alternative to audio-clue-based ones when a camera is available.  The idea of using visual clues for \gls{TSE} is not new\cite{hershey2001audio,rivet2014audiovisual}, although recent neural systems have achieved an impressive level of performance. This is probably because \glspl{NN} can effectively model the relationship between the different modalities learned from a large amount of training data.

Issues and research opportunities remain with the current visual clue-based \gls{TSE} systems. First, most approaches do not consider the speaker tracking problem and assume that the audio and video signals are synchronized. These aspects must be considered when designing and evaluating future \gls{TSE} systems. Second, video processing involves high computational costs, and more research is needed to develop efficient online systems. 

\section{Spatial clue-based TSE}
\label{sec:spatial_tse}
When using a microphone array to record a signal, spatial information can be used to discriminate among sources. In particular, access to multi-channel recordings opens the way to extract target speakers based on their location, i.e., using spatial clues (as indicated in Fig.~\ref{fig:problem}). This section explains how such spatial clues can be obtained and used in \gls{TSE} systems. While enhancing speakers from a given direction has a long research history \cite{Flanagan_85}, we focus here on neural methods that follow the scope of our overview paper.

Note that multi-channel signals can also be utilized in the extraction process using beamforming. Such an extraction process can be used in systems with any type of clue, only requiring that the mixed speech be recorded with multiple microphones. This beamforming process was  \rev{reviewed} in Section~\ref{sec:tse_beamformer}. In this section, we focus specifically on the processing of spatial clues.

\subsection{Obtaining spatial clues}
In some situations, the target speaker’s location is approximately known in advance. For example, for an in-car \gls{ASR}, the driver's position is limited to a certain region in a car. In other scenarios, we might have access to a multi-channel enrollment utterance of the speaker recorded in the same position as the final mixed speech. In such a case, audio source localization methods can be applied. Conventionally, this can be done by methods based on generalized cross-correlation or steered-response power, but recently, deep learning methods have also shown success in this task. An alternative is to skip the explicit estimation of the location and directly extract features in which the location is encoded when a multi-channel enrollment is available. We will detail this approach further in the next section.

Spatial clues can also be obtained from a video using face detection and tracking systems. A previous work \cite{gu2020multi} demonstrated this possibility with a 180-degree wide-angle camera positioned parallel to a linear microphone array\footnote{\url{https://yongxuustc.github.io/grnnbf}}. By identifying the target speaker in the video, the azimuth with respect to the microphone array was roughly approximated. Depth cameras can also be used to estimate not only the azimuth but also the elevation and distance of the speaker.

\subsection{Spatial clue encoder}
\begin{figure*}[tb]
    \centering
    \includegraphics[width=0.8\textwidth]{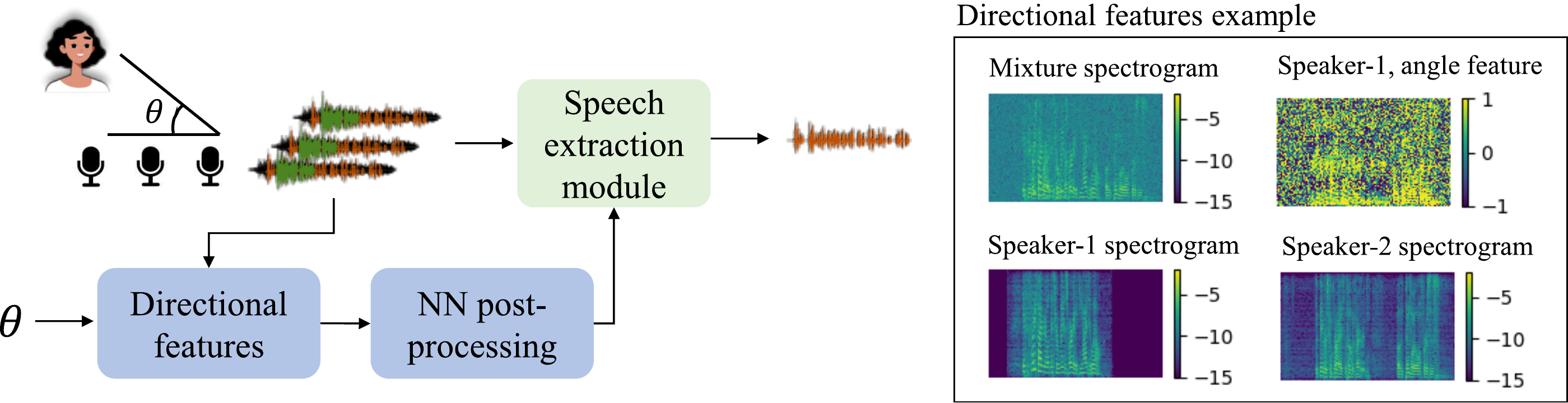}
    \caption{Illustration of usage of spatial clue encoder and directional features}
    \label{fig:spatial_clue}
\end{figure*}

The left part of Fig.~\ref{fig:spatial_clue} shows the overall structure and the usage of a spatial clue encoder, which usually consists of two parts: the extraction of directional features and an \gls{NN} post-processing of them. Two possible forms of spatial clues are dominant in the literature: the angle of the target speaker with respect to the microphone array or a multi-channel enrollment utterance recorded in the target location. Both can be encoded into directional features.

When the spatial clue is \gls{DOA}, the most commonly used directional features are the \textit{angle features}, which are computed as the cosine of the difference between the \gls{IPD} and the \gls{TPD}:
\begin{align}
    \text{AF}[\n,\f] = \sum_{m_1,m_2 \in \mathcal{M}} \cos\biggl( &\text{TPD}\left(m_1,m_2,\phi_s,\f\right) \nonumber \\ 
   & - \text{IPD}\left(m_1,m_2,\n,\f\right)\biggr)
\end{align}
\begin{align}
    &\text{TPD}(m_1,m_2,\phi_s,f) = \frac{2\pi f F_s}{F}\ \frac{\cos \phi_s\ \Delta_{m_1,m_2}}{c} \\
    &\text{IPD}(m_1,m_2,\n,\f) = \angle \mixf^{m_2}[\n,\f] - \angle \mixf^{m_1}[\n,\f], \label{eq:ipd}
\end{align}
where $\mathcal{M}$ is a set of pairs of microphones used to compute the feature, $F_s$ is the sampling frequency, $\phi_s$ is the target direction, $c$ is the sound’s velocity, and $\Delta_{m_1,m_2}$ is the distance from microphone $m_1$ to microphone $m_2$. An example of angle features is shown on the right of Fig.~\ref{fig:spatial_clue}. For time-frequency bins dominated by the source from direction $\phi_s$, the value of the angle feature should be close to 1 or -1. Other directional features have been proposed that exploit a grid of fixed beamformers. A directional power ratio measures the ratio between the power of the response of a beamformer steered into the target direction and the power of the beamformer responses steered into all the directions in the grid. In a similar fashion, a directional signal-to-noise ratio can also be computed, which compares the response of a beamformer in the target direction with the response of a beamformer in the direction with the strongest interference.

If the spatial clue consists of a multi-channel enrollment utterance, the directional feature can be formed as a vector of \gls{IPD}s computed from the enrollment. Alternatively, the \gls{DOA} can be estimated from the enrollment, and the spatial features derived from it can be used.

Note that when using a spatial clue to determine the target speaker, the multi-channel input of the speech extraction module must also be used. This enables the identification of the speaker coming from the target location in the mixture. Furthermore, a target extractor is often implemented as beamforming, as explained in Section~\ref{sec:tse_beamformer}.

\subsection{Combination with other clues}
Although a spatial clue is very informative and generally can lead the \gls{TSE} system to a correct extraction of the target, it does fail in some instances. Estimation errors of \gls{DOA} are harmful to proper extraction. Furthermore, if the spatial separation of the speakers with respect to the microphone array is not significant enough, the spatial clue may not discriminate between them. Combining a spatial clue with audio or visual clues is an option to combat such failure cases. 
    
\subsection{Experimental results}
\begin{figure}
    \centering
    \includegraphics[width=0.5\textwidth]{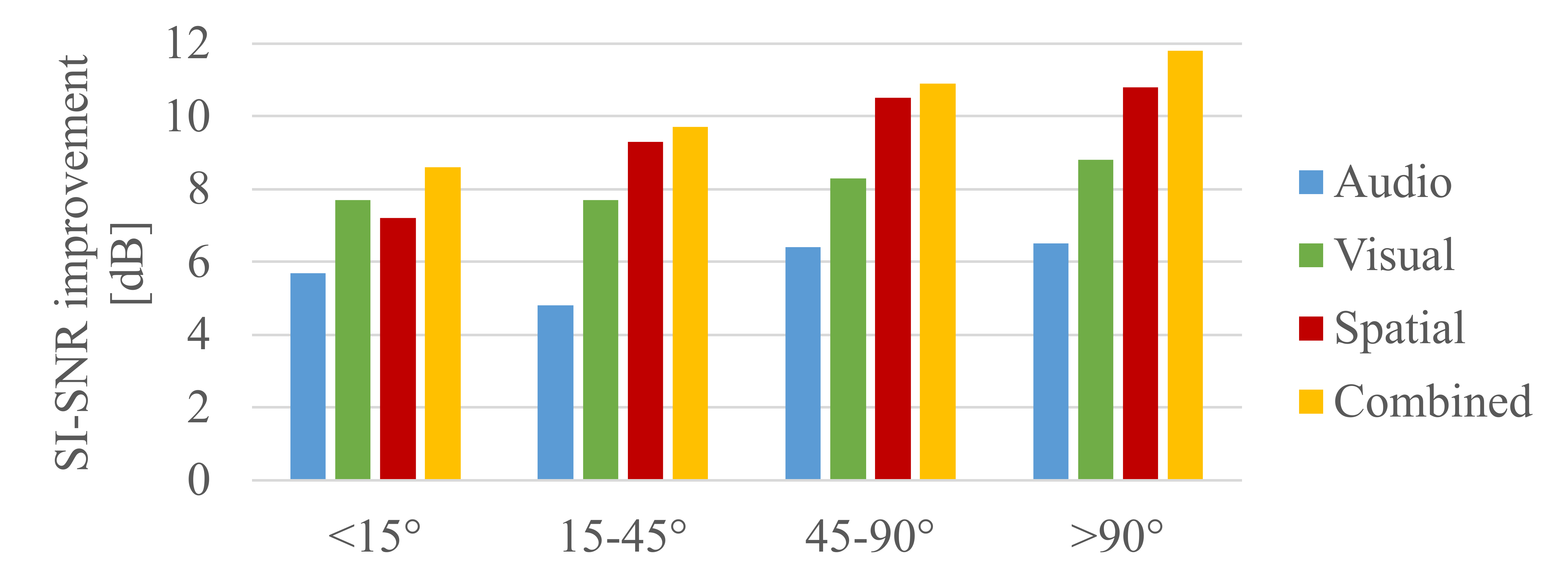}
    \caption{SI-SNR improvement of TSE with audio, visual, and spatial clues in four conditions based on angle separation between speakers \cite{gu2020multi}}
    \label{fig:spatial_res}
\end{figure}

We next report the results from an experiment with spatial clues \cite{gu2020multi} that compared the effectiveness of using audio, visual, and spatial clues. The audio-clue encoder was trained jointly with the extraction module, and the visual encoder was a pre-trained lip-reading network. The target speaker’s direction was encoded in the angle feature. The spatial and visual embeddings were fused with the extraction network by concatenation and the audio embedding with a factorized layer. The extraction module employed a neural network consisting of temporal convolutional layers.

The experiments were performed on a Mandarin audio-visual dataset containing mixtures of two and three speakers. The results in Fig.~\ref{fig:spatial_res} were divided into several conditions, based on the angle separation between the closest speakers. The spatial clue is very effective, although the performance declines when speakers are near each other ($<15\degree$). A combination with other modalities outperformed any individual type of clue in all the conditions.

\subsection{Discussion}
Using spatial clues is a powerful way of conditioning a \gls{TSE} system to extract the target speaker. It relies on the availability of signals from a microphone array and a way to determine the location of the target speaker. Unfortunately, these restrictions limit the applications to some extent. Neural \gls{TSE} methods with spatial clues follow a long history of research on the topic, such as beamforming techniques, and extend them with non-linear processing. This approach unifies the methods with those using other clues and allows a straightforward combination of different clues into one system. Such combinations can alleviate the shortcomings of spatial clues, including the failures when the speakers are located in the same direction from the point of view of the microphones. 

In most current neural \gls{TSE} works, the target speaker’s location is assumed to be fixed. Although the methods should be easily extended to a dynamic case, investigations of such settings remain relatively rare \cite{heitkaemper2019study}.

\section{Extension to other tasks}
\label{sec:extension}

The ideas of \gls{TSE} can be applied to other speech processing tasks, such as \gls{ASR} and diarization.

\subsection{Target-speaker ASR}
\label{ssec:tsasr}
An important application of \gls{TSE} is \gls{TSASR}, where the goal is to transcribe the target speaker's speech and ignore all the interference speakers. The \gls{TSE} approaches we described can be naturally used as a front-end to an \gls{ASR} system to achieve \gls{TSASR}. Such a cascade combination allows for a modular system, which offers ease of development and interpretability. However, the \gls{TSE} system is often optimized with a signal loss, as in Eq.~\eqref{eq:sdr_loss}. Such a \gls{TSE} system inevitably introduces artifacts caused by the remaining interferences, over-suppression, and other non-linear processing distortions. These artifacts limit the expected performance improvement from a \gls{TSE} front-end. 

One approach to mitigate the effect of such artifacts is to optimize the \gls{TSE} front-end with an \gls{ASR} criterion \cite{zmolikova2019speakerbeam}. The \gls{TSE} front-end and the \gls{ASR} back-end are \glspl{NN} and can be interconnected with differentiable operations, such as beamforming and feature extraction. Therefore, a cascade system can be represented with a single computational graph, allowing all parameters to be jointly trained. Such joint-training can significantly improve the \gls{TSASR} performance. 

Another approach inserts a fusion layer into an \gls{ASR} system\cite{delcroix2019end,denisov2019end} to directly perform clue conditioning. These integrated \gls{TSASR} systems avoid any explicit signal extraction step, a decision that reduces the computational cost, although such systems may be less interpretable than cascade systems.

\gls{TSASR} can use the audio clues provided by pre-recorded enrollment utterances\cite{zmolikova2019speakerbeam,denisov2019end,delcroix2019end} or from a keyword (anchor) for a smart-device scenario \cite{king2017robust}, for example. Some works have also exploited visual clues, which can be used for the extraction process and to implement an audio-visual \gls{ASR} back-end, since lip-reading also improves \gls{ASR} performance\cite{Yu2021AudioVisual}.

\subsection{Target-speaker VAD and diarization}
The problem of speech diarization consists of detecting who spoke when in a multi-speaker recording. This technology is essential for achieving, e.g., meeting recognition and analysis systems that can transcribe a discussion between multiple participants. Several works have explored using speaker clues to perform this task \cite{medennikov2020target,ding2019personal}.

For example, a personalized \gls{VAD} \cite{ding2019personal} exploits a speaker embedding vector derived from an enrollment utterance of the target speaker to predict its activity, i.e., whether they are speaking at a given time. In principle, this can be done with a system like that presented in Section \ref{sec:general_framework}, where the output layer performs the binary classification of the speaker activity instead of estimating the target speech signal. Similar systems have also been proposed using visual clues, called audio-visual \gls{VAD}\cite{Sodoyer2006AnAnalysis}. Predicting the target speaker’s activity is arguably a more straightforward task than estimating its speech signal. Consequently, \gls{TSVAD} can use simpler network architectures, leading to more lightweight processing.

The above \gls{TSVAD} systems, which estimate the speech activity of a single target speaker, have been extended to simultaneously output the activity of multiple target speakers\cite{medennikov2020target}. The resulting system achieved the top diarization performance in the CHiME 6 evaluation campaign\footnote{The results of the CHiME 6 challenge can be found at: \url{https://chimechallenge.github.io/chime6/results.html}. The top system used \gls{TSVAD} among other technologies. DiHARD 3 performed a diarization evaluation on the CHiME 6 challenge data. Here the top system also used \gls{TSVAD}:  \url{https://dihardchallenge.github.io/dihard3/results}}.

\section{Remaining issues and outlook}
\label{sec:outlook}
Research toward computational selective hearing has been a long endeavor. Recent developments in \gls{TSE} have enabled identifying and extracting a target speaker's voice in a mixture by exploiting audio, visual, or spatial clues, which is one step closer to solving the cocktail-party problem. Progress in speech processing (speech enhancement, speaker recognition) and image processing (face recognition, lip-reading), combined with deep learning technologies to learn models that can effectively condition processing on auxiliary clues, triggered the progress in the \gls{TSE} field. Some of the works we presented have achieved levels of performance that seemed out-of-reach just a few years ago and are already being deployed in  products\footnote{The following blog details the effort for deploying a visual clue-based \gls{TSE} system for on-device processing: \url{https://ai.googleblog.com/2020/10/audiovisual-speech-enhancement-in.html}.}. 

Despite substantial achievements, many opportunities remain for further research, some of which we list below.

\subsection{Deployment of TSE systems}
Most of the systems we described operate offline and are computationally expensive. They are also evaluated under controlled (mostly simulated mixture) settings. Deploying such systems introduces engineering and research challenges to reduce computational costs while maintaining high performance under less controlled recording conditions. We next discuss some of these aspects.

\subsubsection{Inactive target speaker}
Most \gls{TSE} systems have been evaluated assuming that the target speaker is actively speaking in the mixture. In practice, we may not know beforehand whether the target speaker will be active. We expect that a \gls{TSE} system can output no signal when the target speaker is inactive, which may not actually be the case with most current systems that are not explicitly trained to do so. The inactive target speaker problem is specific to \gls{TSE}. The type of clue used may also greatly impact the difficulty of tackling this problem. For instance, visual voice activity detection \cite{rivet2014audiovisual} might alleviate this issue. However, it is more challenging with audio clues \cite{Zhang2021Towards}, and further research may be required. 

\subsubsection{Training and evaluation criteria}
Most \gls{TSE} systems are trained and evaluated using such signal level metrics as \gls{SNR} or \gls{SDR}. Although these metrics are indicative of the extraction performance, their use presents two issues.

First, they may not always be correlated with human perception and intelligibility or with \gls{ASR} performance. This issue is not specific to \gls{TSE}; it is common to \gls{BSS} and noise reduction methods.
For \gls{ASR} we can train a system end-to-end, as discussed in Section \ref{ssec:tsasr}. When targeting applications for human listeners, the problem can be partly addressed using other metrics for training or evaluation that correlate better with human perception, such as \gls{STOI} or \gls{PESQ}\cite{michelsanti2021overview}. However, controlled listening tests must be conducted to confirm the impact of a \gls{TSE} on human listeners\cite{michelsanti2021overview}.

Second, unlike \gls{BSS} and noise reduction, a \gls{TSE} system needs to identify the target speech, implying other sources of errors. Indeed, failing to identify the target may lead to 
incorrectly estimating an interference speaker or inaccurately outputting the mixture. Although these errors directly impact the \gls{SDR} scores, it would be fruitful to agree on the evaluation metrics that separate extraction and identification performance to better reveal the behavior of \gls{TSE} systems. Signal level metrics might not satisfactorily represent the extraction performance for inactive speaker cases. A better understanding of the failures might help develop \gls{TSE} systems that can recognize when they cannot identify the target speech, which is appealing for practical applications. 

Consequently, developing better training and evaluation criteria are critical research directions.
\subsubsection{Robustness to recording conditions}
Training neural \gls{TSE} systems requires simulated mixtures, as discussed in Section \ref{sec:ssec_training}. Applying these systems to real conditions (multi-speaker mixtures recorded directly with a microphone) requires that the training data match the application scenario relatively well. For example, the type of noise and reverberation may vary significantly depending on where a system is deployed. This raises questions about the robustness of \gls{TSE} systems to various recording conditions. 

Neural \gls{TSE} systems trained with a large amount of simulated data have been shown to generalize to real recording conditions \cite{ephrat2018looking}. However, exploiting real recordings where no reference target speech signal is available could further improve performance. Real recordings might augment the training data or be used to adapt a \gls{TSE} system to a new environment. The issue is defining unsupervised training losses correlated with the extraction performance of the target speech without requiring access to the reference target signal.

Another interesting research direction is combining neural \gls{TSE} systems, which are powerful under matched conditions, with such generative-based approaches as \gls{IVE}\cite{Jansky_20}, which are adaptive to recording conditions. 

\subsubsection{Lightweight and low-latency systems}
Research on lightweight and low-latency \gls{TSE} systems is gaining momentum as the use of teleconferencing systems in noisy environments has risen in response to the Covid pandemic. Other important use cases for \gls{TSE} are hearing aids and hearables, both of which impose very severe constraints in terms of computation costs and latency. The recent DNS\footnote{\url{ https://www.microsoft.com/en-us/research/academic-program/deep-noise-suppression-challenge-icassp-2022/ }} and Clarity\footnote{ \url{https://claritychallenge.github.io/clarity_CC_doc/}} challenges that target teleconferencing and hearing aid application scenarios include tracks where target speaker clues (enrollment data) can be exploited. This demonstrates the growing interest in practical solutions for \gls{TSE}.

Since \gls{TSE} is related to \gls{BSS} and noise reduction, the development of online and low-latency \gls{TSE} systems can be inspired from the progress of \gls{BSS}/noise reduction in that direction. However, \gls{TSE} must also identify the target speech, which may need specific solutions that exploit the long-context of the mixture to reliably and efficiently capture a speaker’s identity.

\subsubsection{Spatial rendering}
For applications of \gls{TSE} to hearing aids or hearables, sounds must be localized in space after the \gls{TSE} processing. Therefore, a \gls{TSE} system must not only extract the target speech but also estimate its direction to allow rendering it so that a listener feels the correct direction of the source.

\subsection{Self-supervised and cross-modal learning}
A \gls{TSE} system identifies the target speech in a mixture based on the intermediate representation of the mixture and the clue. Naturally, \gls{TSE} benefits from better intermediate representations. For example, speech models learned with self-supervised learning criteria have gained attention as a way to obtain robust speech representations. They have shown potential for pre-training many speech processing downstream tasks, such as \gls{ASR}, speaker identification, and \gls{BSS}. Such self-supervised models could also reveal advantages for \gls{TSE} since they could improve robustness by allowing efficient pre-training on various acoustic conditions. Moreover, for audio-based \gls{TSE}, using the same self-supervised pre-trained model for the audio clue encoder and the speech extraction module will help to learn the common embedding space between the enrollment and speech signals in the mixture. Similarly, the progress in cross-modal learning, which aims to learn the joint representation of data across modalities, could benefit  such multi-modal approaches as visual clue-based \gls{TSE}.

\subsection{Exploring other clues}
We presented three types of clues that have been widely used for \gls{TSE}. However, other clues can also be considered. 
\rev{For example, recent works have explored other types of spatial clues such as the distance \cite{tzinis22_interspeech}. }
\rev{Moreover}, humans do not only rely on physical clues to perform selective hearing. We also use more abstract clues, such as semantic ones.
Indeed, we can rapidly focus our attention on a speaker when we hear our name or a topic we are interested in. Reproducing a similar mechanism would require \gls{TSE} systems that \rev{operate} with semantic clues, which introduces novel challenges concerning how to represent semantic information and exploit it within a \gls{TSE} system. 
\rev{Some works have started to explore this direction, such as conditioning on languages\cite{Borsdorf_2021} or more abstract concepts \cite{Ohishi_2022}.}

\rev{Other interesting clues consist} of signals that measure a listener's brain activity to guide the extraction process. Indeed, the \gls{EEG} signal of a listener focusing on a speaker correlates with the envelope of that speaker’s speech signal. Ceolini et al. identified the possibility of using \gls{EEG} as clues for \gls{TSE} with a system similar to the one described in Section \ref{sec:general_framework}\cite{ceolini2020brain}. An \gls{EEG}-guided \gls{TSE} might open the door for futuristic hearing aids controlled by the user's brain activity, which might automatically emphasize the speaker a user wants to hear. However, research is still needed because developing a system that requires marginal tuning to the listener is especially challenging. Moreover, collecting a large amount of training data is very complicated since it is more difficult to control the quality of such clues. Compared to audio and visual \gls{TSE} clues, \gls{EEG} signals are very noisy and affected by changes in the attention of the listener, body movements, and other factors.

\subsection{Beyond speech}

Human selective listening abilities go beyond speech signals. For example, we can focus on listening to the part of an instrument in an orchestra or switch our attention to a siren or a barking dog. In this paper, we focused on \gls{TSE}, but similar extraction problems have also been explored for other audio-processing tasks. For example, much research has been performed on extracting the track of an instrument in a piece of music \rev{conditioned on, e.g., the type of instrument \cite{Seetharaman_2019}, video of the musician playing \cite{Zhao_2018_ECCV}, or \gls{EEG} signal of the listener \cite{Cantisani_2021}. These approaches may be important to realize, e.g., audio-visual music analysis\cite{Duan_2019}.} 

Recently, the problem was extended to the extraction of arbitrary sounds from a mixture\rev{\cite{ochiai2020listen,gfeller2020one}}, e.g., extracting the sound of a siren or a klaxon from a recording of a mixture of street sounds. We can use such systems as that introduced in Section \ref{sec:general_framework} to tackle these problems, where the clue can be a class label indicating the type of target sound\rev{\cite{ochiai2020listen}}, the enrollment audio of a similar target sound \rev{\cite{gfeller2020one}}, a video of the sound source\rev{\cite{owens2018audio}} \rev{or a text description of the target sound\cite{liu22w_interspeech}}.  \rev{Target sound extraction may become an important technology to design, e.g., hearables or hearing aids that could filter out nuisances and emphasize important sounds in our surroundings, or audio visual scene analysis \cite{owens2018audio}}.

Psycho-acoustic studies suggest that humans process speech and music partly using shared auditory mechanisms and that exposure to music can lead to better discrimination of speech sounds \rev{\cite{asaridou2013speech}}. It would be interesting to explore whether, similarly to humans, \gls{TSE} systems could benefit from exposure to other acoustic signals by training a system to extract target speech, music, or arbitrary sounds.

\section{Resources}
\label{sec:resources}
We conclude by providing pointers to selected datasets and toolkits available for those motivated to experiment with \gls{TSE}. 

\gls{TSE} works mostly use datasets designed for \gls{BSS}. These datasets consist generally of artificial mixtures generated from the isolated signals of the individual speakers and background. This allows evaluation of the performance by comparing the estimated signals to the original references. Additionally, \gls{TSE} methods also require a clue, i.e., an enrollment utterance for the target speaker or video signal. We can obtain enrollment utterances by choosing a random utterance of the target speaker from the same database, provided that the utterance is different from the one in the mixture. For a video clue, it requires using an audio-visual dataset.
The top of Table \ref{tab:resources} lists some of the most commonly used datasets for audio and visual \gls{TSE}.

Several implementations of \gls{TSE} systems are openly available and listed in the lower part of Table \ref{tab:resources}. Although there are no public implementations for some of the visual \gls{TSE} systems, they can be re-implemented following the audio \gls{TSE} toolkits and using openly available visual feature extractors such as FaceNet, which was used in some previous works \cite{ephrat2018looking,ochiai2019multimodal,sato2021multimodal}.

\begin{table*}[tb]
    \centering
    \caption{Some datasets and toolkits}
    \begin{tabular}{@{}l@{ }l@{ }l@{ }l@{}}
    \toprule
              & Name & Description & Link  \\
       \midrule
    \multirow{8}{*}{\rotatebox[origin=c]{90}{Dataset}}       
    & WSJ0-mix & Mixtures of two or three speakers & \url{www.merl.com/demos/deep-clustering} \\
    & WHAM(\rev{R})\rev{!} & Noisy and reverberant versions of WSJ0-mix &  \url{wham.whisper.ai}\\
    & Librimix & Larger dataset of mixtures of two or three speakers & \url{github.com/JorisCos/LibriMix}\\
    &LibriCSS & Meeting-like mixtures recorded in a room &\url{github.com/chenzhuo1011/libri_css} \\
    & MC-WSJ0-mix & Spatialized version of WSJ0-2mix & \url{www.merl.com/demos/deep-clustering} \\
    & SMS-WSJ & Multi-channel corpus based on WSJ & \url{github.com/fgnt/sms_wsj} \\
    & LRS & Audio-visual corpus from TED or BBC videos &\url{www.robots.ox.ac.uk/~vgg/data/lip_reading} \\
    & AVSpeech & Very large audio-visual corpus from YouTube videos &\url{looking-to-listen.github.io/avspeech} \\
    \midrule
    \multirow{6}{*}{\rotatebox[origin=c]{90}{Tools}}  
    & SpeakerBeam & Time-domain audio-based \gls{TSE} system & \url{github.com/butspeechfit/speakerbeam} \\
    &SpEx+      & Time-domain audio-based \gls{TSE} system \cite{ge2020spex} & \url{github.com/xuchenglin28/speaker_extraction_SpEx} \\
    & VoiceFilter & Time-domain audio-based \gls{TSE} system (Unofficial) \cite{wang2018voicefilter}& \url{github.com/mindslab-ai/voicefilter} \\
    & Multisensory & Visual clue-based \gls{TSE}  \cite{owens2018audio}& \url{github.com/andrewowens/multisensory} \\
    & AV Speech enh.& Face landmark-based visual clue-based \gls{TSE} \cite{morrone2019face}& \url{github.com/dr-pato/audio_visual_speech_enhancement} \\
    & FaceNet & Visual feature extractor used in \cite{ephrat2018looking,ochiai2019multimodal,sato2021multimodal} & \url{github.com/davidsandberg/facenet} \\
    \bottomrule
    \end{tabular}
    \label{tab:resources}
\end{table*}

\section{Acknowledgments}
This work was partly supported by the Czech Ministry of Education, Youth and Sports from project no. LTAIN19087 "Multi-linguality in speech technologies." Computing on IT4I supercomputer was supported by the Czech Ministry of Education, Youth and Sports from the Large Infrastructures for Research, Experimental Development and Innovations project "e-Infrastructure CZ – LM2018140".
The figures contain elements designed by pikisuperstar/Freepik. 

%


\bibliographystyle{ieeetr}
\bibliography{references}

\end{document}